\documentclass[reprint,showpacs,prd,preprintnumbers,amssymb,nofootinbib,superscriptaddress,aps]{revtex4-1}
\pdfoutput=1
\usepackage{amsmath,amssymb,bm,natbib}
\usepackage{bbold}
\usepackage{epsfig}
\usepackage{hyperref}
\usepackage{graphicx}
\usepackage{csquotes}
\usepackage{mathtools}
\usepackage{slashed}
\usepackage{color,colordvi}
\usepackage{soul}
\usepackage{xcolor}
\usepackage{booktabs,multirow,array}
\usepackage{multirow}
\usepackage{tabularx,ragged2e,booktabs,caption}
\usepackage{comment}
\usepackage{slashed}
\usepackage{float}
\usepackage[normalem]{ulem}
\renewcommand{\(}{\left(}
\renewcommand{\)}{\right)}

\renewcommand{\}}{\right\rbrace}

\newcommand{\nn}{\nonumber}

\newcommand{\order}[1]{\mathcal{O}\({#1}\)}
\newcommand{\alphas}{\alpha_\mathrm{s}}

\newcommand{\GeV}{\,\mathrm{GeV}}

\newcommand{\MeV}{\,\mathrm{MeV}}

\begin{document}
	\title{The QCD Static Energy using Optimal Renormalization and Asymptotic Pad\'e-approximant Methods}
	\author{B. Ananthanarayan}
	\email{anant@iisc.ac.in}
	\affiliation{Centre for High Energy Physics, Indian Institute of Science, Bangalore 560 012, India}
	\author{Diganta Das}
	\email{diganta99@gmail.com}
	\affiliation{Department of Physics and Astrophysics, University of Delhi, Delhi 110007, India}
	\author{M. S. A. Alam Khan}
	\email{alam.khan1909@gmail.com}
	\affiliation{Centre for High Energy Physics, Indian Institute of Science, Bangalore 560 012, India}
	\begin{abstract}
The perturbative QCD static potential and ultrasoft contributions, which together give the static energy, have been calculated to three- and four-loop order respectively, by several authors. Using the renormalization group, and Pad\'e approximants, we estimate the four-loop corrections to the static energy. We also employ the optimal renormalization method and resum the logarithms of the perturbative series in order to reduce sensitivity to the renormalization scale in momentum space. This is the first application of the method to results at these orders. The convergence behaviour of the perturbative series is also improved in position space using the Restricted Fourier Transform scheme. Using optimal renormalization, we have extracted the value of $\Lambda^{\overline{\textrm{MS}}}_{\textrm{QCD}}$ at different scales for two active flavours by matching to the static energy from lattice QCD simulations.
	\end{abstract}
	\maketitle
	\section{Introduction}
The static potential energy of quantum chromodynamics~(QCD) is the non-Abelian analogue of the well-known Coulomb potential energy of quantum electrodynamics. The short distance part of this quantity is calculated in non-relativistic QCD (NRQCD)\cite{Caswell:1985ui,Bodwin:1994jh} framework and involves the evaluation of Feynman diagrams. It has been studied extensively in recent years and analytical results are known to three-loop. At four-loop order contributions involving the ultrasoft gluons that start to contribute from three-loop order in perturbation series are known.  Such contributions only appear for short distances when system is weakly coupled. They are calculated using weakly coupled potential non-relativistic QCD (pNRQCD)\cite{Pineda:1997bj,Brambilla:2004jw} formalism. Hence, we discuss only weakly coupled regime of pNRQCD in this article. The static potential depends on the renormalization scale $\mu$, the ultrasoft factorization scale $\mu_{\rm us}$, and magnitude of the three momentum transfer between the heavy sources $p\left(=|\mathbf{p}|\right)$. In this article $\overline{\text{MS}}$ renormalization scheme is used.\par
The first attempt to perform the full three-loop (numerical) calculations was by two independent groups \cite{Anzai:2009tm,Smirnov:2009fh}. 
The results were found to be in agreement. Analytical calculations were presented almost six years later in ref~\cite{Lee:2016cgz}. At three-loop order, ultrasoft gluons also appear which are capable of changing singlet to octet state of the system and \textit{vice-versa}. Such contributions were first pointed in ref~\cite{Appelquist:1977es}, and were calculated in ref~\cite{Brambilla:1999qa} in pNRQCD. The renormalization group (RG) improvement of the ultrasoft terms at three-loop order was first discussed in ref~\cite{Pineda:2000gza}. The next order calculations for the ultrasoft terms can be found in ref~\cite{Brambilla:2006wp} and their resummation is discussed in ref~\cite{Brambilla:2009bi}.

QCD is known to be non-perturbative at long distance, and this fact is manifest in lattice QCD (LQCD) simulations. The static energy, that includes the static potential and the ultrasoft contributions, can also be computed in LQCD simulations, and hence it becomes a topic of great interest to extract various parameters of the theory. Some recent LQCD simulation results for the static energy can be found in refs~\cite{Bazavov:2014soa,Karbstein:2018mzo,Karbstein:2014bsa,Takaura:2018vcy,Takaura:2018lpw,Bazavov:2019qoo}, and they support the Cornell potential type behaviour for the heavy quark-antiquark system.

The potential energy between a heavy quark and anti-quark pair, is an important ingredient to describe, among other things, non-relativistic bound states like quarkonia~\cite{Brambilla:1999xf,Brambilla:2004jw}, quark masses~\cite{Hoang:2000fm,Pineda:2001zq,Beneke:2014pta,Penin:2014zaa,Ayala:2014yxa,Kiyo:2015ufa,Kiyo:2015ooa,Beneke:2015zqa,Mateu:2017hlz,Peset:2018ria} and threshold production of top quarks~\cite{Hoang:2013uda,Beneke:2013jia,Beneke:2015kwa} etc.

In this article, we have addressed the following issues:
\begin{itemize}
\item At four-loop order, the constant term contributing to the perturbative series is as yet unknown. It requires calculation of four-loop Feynman diagram. In the absence of such a calculations we use RGE and Pad\'e approximants~\cite{Chishtie:2001mf} to get the estimate for the four-loop coefficients. 
\item The RG improvement of the static energy by resummation of all RG-accessible running logarithms following the method advocated in refs~\cite{Maxwell:1999dv,Maxwell:2000mm,Maxwell:2001he,Ahmady:1999xg,Ahmady:2002fd,Ahmady:2002pa,Ananthanarayan:2016kll,Abbas:2012py}, and for the first time applied in this paper to the ultrasoft logarithms present in the static energy. All the leading and next-to-leading logarithms at each order in perturbation theory that can be accessed through the RG equation (RGE) are called RG-accessible logarithms. We call this renormalization group improved perturbative series as RG-summed or optimal renormalized series. It should be noted that our resummation scheme discuss about RG-improvement with respect to renormalization scale whereas ref~\cite{Pineda:2000gza,Brambilla:2009bi} deals with the RG-improvement with respect to ultrasoft scale.
\item The convergence of the static energy is improved to the four-loop order in position space using the Restricted Fourier Transform (RFT) proposed in ref~\cite{Karbstein:2013zxa}.
\item We use the RG-summed and unsummed forms of the static energy in momentum space to extract $\Lambda^{\overline{\textrm{MS}}}_{\textrm{QCD}}$ to four-loop from the LQCD inputs from ref~\cite{Karbstein:2018mzo}.
\item In comparison to the unsummed series, the RG-summed static energy in momentum space gives better fit to the static energy obtained from LQCD in ref~ \cite{Karbstein:2018mzo}.
\item The four-loop RG-summed static energy is used as a trial case to show that these improvements (scale sensitivity and better fit) also persist at the higher order.
\end{itemize}

The scheme of this paper is as follows: In section~\ref{sec:pert_pot}, we discuss the perturbative treatment to the static energy and various pieces associated with this quantity in the weak coupling limit. In section~\ref{sec:RG_ser}, we calculate some contributions to the static energy at the four-loop using the RGE. In section~\ref{sec:Pade_est}, we use Pad\'e approximants to estimate all the four-loop coefficients even the one that is not accessible with the RGE. Some of the estimates are found in agreement with RGE solutions. In section~\ref{sec:RG_sum}, we perform all order RG-summation of certain running logarithms and show that this RG-improvement will bring down the sensitivity to the renormalization scale. In section~\ref{sec:RFT}, we discuss the improvement of the static energy to the four-loop order in position space by removing the pathological uncontrolled contributions using RFT. The inputs from the previous sections are applied in section~\ref{sec:lat_input} to fit the LQCD inputs for the static energy to extract the $\Lambda^{\overline{\textrm{MS}}}_{\textrm{QCD}}$ for two active flavours in momentum space. A discussion is presented in section \ref{sec:conc} and we summarize our results in  section~\ref{sec:summary}. Appendix~\ref{app:QCDbeta} contains the QCD beta function coefficients. Appendix~\ref{app:alpha_run} contains the formula used in for running of the strong coupling constant with momentum and also in terms of $\Lambda^{\overline{\textrm{MS}}}_{\textrm{QCD}}$. Appendix~\ref{app: loop_coef} contains the known contribution to the static energy. Appendix~\ref{app:Vr} contains the useful formula for calculating the restricted and unrestricted versions of position space static potential and the static energy. Appendix~\ref{app:resVr} contains the final result of uncontrolled contribution to the static energy in position space to the four-loop order. 
\section{The Perturbative QCD-Static Energy \label{sec:pert_pot}}
A heavy quark and anti-quark system, with heavy quark mass $m_Q$ and relative velocity $v<1$, is non-relativistic in nature and various scales \cite{Brambilla:2004jw,Beneke:1997zp} present in the system are hard scale $\sim \mathcal{O}(m_Q)$, soft scale $\sim\mathcal{O}(m_Q v)$, and ultrasoft scale $\sim\mathcal{O}(m_Q v^2)$. If these scales are well separated then we can integrate them one by one to study only relevant degrees of freedom. Integrating out the hard scale from QCD gives the non-relativistic QCD (NRQCD) in which the soft and the ultrasoft degrees of freedom are dynamical. Contributions to the static energy at different orders were calculated using this framework. The one-loop perturbative calculations for massless quarks were first performed in the late 1970s and can be found in refs~\cite{Appelquist:1977es,Appelquist:1977tw,Susskind,Fischler:1977yf} and in the massive case in ref~\cite{Billoire:1979ih}. Two-loop massless calculations appeared in refs~\cite{Melles:1998dj,Schroder:1998vy, peter:1998ml, Peter:1997me} while the massive case were considered in refs~\cite{Melles:2000ml,Melles:2000ey, sumino:2002ms, Hoang:2000fm}.\par
The pNRQCD is obtained from NRQCD by integrating out the soft scale and this formalism is best suitable for studying the threshold systems. The heavy quark and anti-quark systems in this formalism are described by color singlet fields $S$ and color octet fields $O$. The gauge fields are multipole expanded about the inter-quark separation and the gauge invariant Lagrangian for pNRQCD\cite{Brambilla:1999xf,Brambilla:2004jw}, at leading order in $1/m_Q$ is given by
\begin{align}
\mathcal{L}_{\text {pNRQCD }}=& \mathcal{L}_{\text {light}}+\operatorname{Tr}\left(\mathrm{S}^{\dagger}\left(i \partial_{0}-V_{s}(r,\mu_{us})+\ldots\right) \mathrm{S}\right)\nn\\&+\operatorname{Tr}\left(\mathrm{O}^{\dagger}\left(i D_{0}-V_{o}(r,\mu_{us})+\ldots\right) \mathrm{O}\right)\nonumber \\ &+g V_{A}(r,\mu_{us}) \operatorname{Tr} \left(\mathrm{O}^{\dagger} \boldsymbol{r} \cdot \mathbf{E} \mathrm{S}+\mathrm{S}^{\dagger} \boldsymbol{r} \cdot \mathbf{E} \mathrm{O}\right) \nn\\&+g \frac{V_{B}(r,\mu_{us})}{2} \operatorname{Tr}\left(\mathrm{O}^{\dagger} \boldsymbol{r} \cdot \mathbf{E} \mathrm{O}+\mathrm{O}^{\dagger} \mathbf{O} \boldsymbol{r} \cdot \mathbf{E}\right)\nn\\&-\frac{1}{4} F_{\mu \nu}^{a} F^{\mu \nu a}\, ,
\label{eq:lag_pnrqcd}
\end{align}
where $\mathcal{L}_{\text {light}}$ is Lagrangian for light quarks, $iD_0 O \equiv i \partial_0 O-g \left[A_0(\mathbf{R},t),O\right]$ and $\mathbf{E}$ are chromoelectric field strength. $V_s(r,\mu_{us})$ and $V_o(r,\mu_{us})$ are singlet and octet potential at leading order in $r$ while $V_A$ and $V_B$ appear at sub-leading order in $r$. It should be noted that these potential will also depend on the renormalization scale $\mu$ for any finite order calculation. The potentials appearing at higher order in $1/m_Q$ are hidden in ellipses which disappear in static limit $m_Q\rightarrow\infty$. The gauge fields are already multipole expanded about inter-quark separation ($\mathbf{R}$) in eq(\ref{eq:lag_pnrqcd}) and therefore, $F^{\mu \nu a}\equiv F^{\mu \nu a}(\mathbf{R},t)$.

In the weak coupling regime, the system has the hierarchy of scales:	
\begin{align}
m_Q\gg m_Q~v\gg m_Q~v^2\gg\Lambda^{\overline{\textrm{MS}}}_{\textrm{QCD}},
\label{wc_limit}
\end{align}
and in this limit, the static energy can be written as a perturbative series in the strong coupling constant $\alpha_{s}$. The singlet static energy to known orders can be written as:
	\begin{align}
	E_{\mathrm{0}}(p,\mu)=V_s(p,\mu,\mu_{\mathrm{us}})+\delta^{\rm us}(p,\mu,\mu_{\mathrm{us}})\, ,
	\label{E0}
	\end{align}
where $V_s(p,\mu,\mu_{\mathrm{us}})$ is the perturbative static potential  and $\delta^{\rm us}(p,\mu,\mu_{\mathrm{us}})$ are contribution from ultrasoft gluons. The static potential encodes the interaction of the quarks and gluons degrees of freedom in the singlet state and is known to three loop. It takes the following form:
\begin{align}
&V_s(p,\mu,\mu_{us})=\frac{-4 \pi^2 C_F}{p^2}\nonumber\\&\times\sum _{i=0}^n \sum _{j=0}^i x^{i+1} L^j \left(T_{i,j,0}+\sum _{k=1}^{i-2} \theta (i-3) \widetilde{T}_{i,j,k}^{\rm us} \log ^k\left(\frac{\mu _{us}^2}{p^2}\right)\right)\nonumber\\&\hspace{2cm}+\order{x^{n+2}}\,.
\label{Vpert}
\end{align}
In the above, $L\equiv \log\left(\frac{\mu^2}{p^2}\right)$ is running logarithm, and $C_F=4/3$ is color factor of the $SU(3)$ representation. The expansion parameter in the above equation is defined as $x\equiv(\alpha_s(\mu)/\pi)$, and any argument of $x$ indicates the scale at which it is evaluated. The coefficients $T_{i,0,0}$ are the $ i^{th}-$loop perturbative contributions. The one-loop coefficient $T_{1,0,0}$ can be found in refs~\cite{Billoire:1979ih,Fischler:1977yf}, the two-loop coefficient $T_{2,0,0}$ in refs~\cite{Schroder:1998vy,peter:1998ml,Peter:1997me} and the three-loop coefficient $T_{3,0,0}$ in refs~\cite{Smirnov:2009fh,Anzai:2009tm,Lee:2016cgz}. All these coefficients are collected in appendix~\ref{app: loop_coef}. The coefficients of infrared logarithms, $\widetilde{T}_{i,j,k}^{\rm us}$, in the static potential and can be found in refs~ \cite{Brambilla:1999qa,Brambilla:2006wp}.\par
In effective field theory language, the static potential is a matching coefficient which depends upon the factorization scale $\mu_{\rm us}$. The presence of the logarithmic terms in the static potential at three-loop were first pointed out in the ref~\cite{Appelquist:1977es} and they act as a source of infrared divergences. The ultrasoft part $\delta^{\rm us}$ is now known to next to the leading order (NLO)~\cite{Brambilla:1999qa,Brambilla:2006wp} but it contributes to the static energy from three-loop order(briefly discussed in section~\ref{sec:RFT}). They carry the information of the dynamical ultrasoft gluon degrees of freedom with the ultraviolet cut-off $\mu_{\rm us}$. This scale acts as a source of ultra-violet divergences for these gluons. Both divergences, however, cancel with each other for the static energy which results in non-analytic dependence $\left(\sim \alpha_{s}^n \log^m\left(\alpha_{s}\right)\right)$ in terms of the expansion parameter and the total energy in eq(\ref{E0}) takes the form:
\begin{align}
E_0(p,\mu)&=-\frac{4 \pi^2 C_F }{p^2} \sum _{i=0}^n \sum _{k=0}^{\tiny{\substack{(i-2)\\ \times\theta (i-3)}}} x(p)^{i+1} T_{i,0,k}\log ^k(x(p))\nonumber\\&\hspace{4cm}+\order{x(p)^{n+2}}\nonumber\\& \equiv -\frac{4 \pi^2 C_F }{p^2} W(x(p),L=0)\, ,
\label{E01}
\end{align}
where $T_{i,0,0}\overset{i>2}{=} T^{\rm pert}_{i,0,0}+\delta T^{\rm us}_{i,0,0}$ and constant terms ($\delta T^{\rm us}_{i,0,0}$) are the contributions from the ultrasoft gluons and can be found in ref~\cite{Brambilla:2006wp}.
	
The ultrasoft contributions to the static energy are now known to the four-loop order, but the perturbative correction to static potential are yet to be calculated at this order. Some of the contributions to the static energy at this order can be accessed using RGE and they are discussed in the next section~\ref{sec:RG_ser}.
	
\section{RG Solutions of the four-loop contributions\label{sec:RG_ser}}
We can rewrite eq(\ref{E01}) in terms of the coupling at the renormalization scale $\mu$ using eq(\ref{alphasmu}) as:
	\begin{align}
	E_0(p,\mu)&=-\frac{\left(4 \pi^2 C_F \right) }{p^2} \nonumber\\&\times \sum _{i=0}^n \sum _{j=0}^i \sum _{k=0}^{\tiny{\substack{(i-j-2)\\ \times \theta (i-j-3)}}} x^{i+1} L^j \log ^k(x) T_{i,j,k}\nonumber\\&=-\frac{4 \pi^2 C_F }{p^2} W(x,L)\, .
	\label{E0exp}
	\end{align}
Despite the explicit dependence, all-orders series should be independent of the renormalization scale $\mu$. Mathematically, this implies that any perturbative series $W(x,L)$ must satisfy the RGE:
	\begin{equation}
	\mu^2 \frac{d^2}{d\mu^2} W(x, L) = 0\, ,
	\label{RG_W}
	\end{equation} 
	where
	\begin{equation}
	\mu^2 \frac{d^2}{d\mu^2} = \frac{\partial }{\partial L} + \beta(x) \frac{\partial }{\partial x} \, .
	\label{RGS}
	\end{equation}
The QCD beta function $\beta$ is defined in terms of $x$ as:
	\begin{equation}
	\beta(x) = \mu^2 \frac{d^2}{d\mu^2} x = -\sum_{i=0}^{\infty} \beta_i x^{i+2}\, ,
	\label{beta}
	\end{equation}	
where the coefficients $\beta_i$ up to five-loop order~\cite{Gross:1973id, Caswell:1974gg,Jones:1974mm,Tarasov:1980au,Larin:1993tp,vanRitbergen:1997va,Czakon:2004bu,Baikov:2016tgj,Herzog:2017ohr} are given in appendix~\ref{app:QCDbeta}. The RGE can be used to solve iteratively for the RG-accessible terms in terms of the QCD $\beta_i$ coefficients and the lower order RG-inaccessible coefficients $T_{i,0,k}$. 

The RGE for the $W(x,L)$ along with known results to three-loop and QCD beta functions allow us to extract the RG-accessible four-loop coefficients $T_{4,1,0}$, $T_{4,2,0}$, $T_{4,3,0},T_{4,1,1}$ and $T_{4,4,0}$. To obtain them, we note that 
	\begin{align}
	&\left(\frac{\partial }{\partial L}+\beta(x) \frac{\partial}{\partial x}\right) W(x,L) = \left( T_{1,1,0}-\beta_0\right)x^2+(T_{2,1,0}-\beta_1 \nonumber \\ & - 2 \beta_0 T_{1,0,0})x^3+ \left(2T_{2,2,0}-2\beta_0 T_{1,1,0}\right)x^3 L+(T_{3,1,0}-\beta_2\nonumber \\ &-2\beta_1 T_{1,0,0}-3\beta_0 T_{2,0,0})x^4+\big(-3\beta_{1} T_{2,1,0}-2\beta_1 T_{1,1,0}\nonumber \\ &+2T_{3,2,0}\big)x^4 L+ \left(3T_{3,3,0}-3\beta_{1} T_{2,2,0}\right)x^4 L^2+\big(T_{4,1,0}-\beta_3\nonumber \\ &-2\beta_2 T_{1,0,0}-3\beta_1 T_{2,0,0}-4\beta_{1} T_{3,0,0}-\beta _0 T_{3,0,1}+\log (x) T_{4,1,1}\nonumber \\ &-4 \beta _0 \log (x) T_{3,0,1}\big)x^5 +(2 T_{4,2,0}-2 T_{1,1,0} \beta _2-3 T_{2,1,0} \beta _1\nonumber\\&-4 \beta _0 T_{3,1,0})x^5 L+(3T_{4,3,0}-3\beta_1 T_{2,2,0}-4\beta_{1} T_{3,2,0})x^5 L^2\nonumber \\ &+\left(4T_{4,4,0}-4\beta_{0} T_{3,3,0}\right)x^5 L^3+ \order{x^6} =0\, .
	\label{eq:rg_W}
	\end{align} 
Coefficients of $x^i L^j \log^k(x)$ of the above equation gives the RG-accessible coefficients
	\begin{align}
	T_{4,1,0}&= 4 \beta_0 T_{3,0,0}\beta_0 +\beta _0 T_{3,0,1}+3 \beta_1 T_{2,0,0}+2 \beta_2 T_{1,0,0}+\beta_3 \, ,\nonumber\\
	T_{4,2,0}&=6 \beta_0^2 T_{2,0,0}+7 \beta_1 \beta_0 T_{1,0,0}+3 \beta_2 \beta_0+\frac{3 \beta_1^2}{2} \, ,\nn \\
	T_{4,3,0}&=4 \beta_0^3 T_{1,0,0}+\frac{13}{3} \beta_1 \beta_0^2\, ,\label{RG_sol}\\
	T_{4,4,0}&=\beta_0^4\, ,\nonumber\\
	T_{4,1,1}&=4 \beta _0 T_{3,0,1}\, \nn.
	\end{align}
The coefficients $T_{i,0,k}$ can not be obtained using RGE and are known as the RG-inaccessible terms. The calculation of such terms involves the evaluation of all the Feynman diagrams relevant for that order. Such calculation are yet to be performed so we use asymptotic Pad\'e approximant (APAP) to estimate unknown coefficient, $T_{4,0,0}$, which is discussed in the next section. 
	
\section{Pad\'e estimate of the four-loop contribution \label{sec:Pade_est}}
Pad\'e approximants are rational functions that can be used to estimate the higher order terms of a series from its lower order coefficients. Both the original series and the Pad\'e approximants have the same Taylor expansion to a given order, the next term is taken as its prediction. They are used in refs~\cite{Samuel:1995jc,Ellis:1996zn,Ellis:1997sb,Elias:1998bi, Chishtie:1998rz,Ahmady:1999xg, Elias:2000iw,Chishtie:2000ex}  in the past to improve the higher order perturbative results for QCD.  It is worth to mention that they were first used for the static potential in ref~\cite{Chishtie:2001mf} and we are extending the results using the same method for static energy to the four-loop order. The procedure used is explained in this section.

Note that the ultrasoft corrections to the static energy are already known to the four-loop order, so we do not need to predict them using Pad\'e approximants. They can be added to the predicted values at the end of calculations. We rewrite the perturbative series in eq(\ref{E0exp}), without the ultrasoft corrections as $\widetilde{W}(x,L)$, in the following form:
	\begin{equation}\label{eq:Wxl2}
	\widetilde{W}(x,L)= 1 + R_1 x + R_2 x^2 + R_3 x^3 + R_4 x^4 +\cdots + R_N x^N + \cdots\, ,
	\end{equation}
where \begin{equation}\label{eq:Rs}
		R_i\equiv\sum _{j=0}^i L^j T_{i,j,0}\, .
	\end{equation}
In general, if the series coefficients $\{R_1, R_2, R_3, \cdots, R_N \}$ are known then the Pad\'e approximant for this series is denoted by $\widetilde{W}^{[N-M|M]}$, and given as:
	\begin{equation}
	\widetilde{W}^{[N-M|M]} = \frac{1+A_1 x+A_2 x^2+\cdots+A_{N-M} x^{N-M}}{1+B_1 x+\cdots+B_M x^{M}}\, ,
	\end{equation}
can be used to estimate $R_{N+1}$. For example, if only the NLO term of eq(\ref{eq:Wxl2}), $R_1$, is known then the next coefficient, $R_2$, is estimated from $\widetilde{W}^{[0|1]}$ by Taylor expanding it for small $x$ as
	 \begin{equation}
	 \widetilde{W}^{[0|1]} = \frac{1}{1-R_1 x} = 1 + R_1 x + R_1^2 x^2 + \order{x^3}\, ,
	 \end{equation}
	 {\emph i.e.,} $R_1^2$ is the Pad\'e approximant prediction for $R_2$.\par
For the static energy, the coefficients $R_1, R_2, R_3$ are known and can be read off by comparing the $x^i$-th terms in eq(\ref{E0exp}) with eq(\ref{eq:Wxl2}). To predict the unknown four-loop coefficient $R_4$ in eq(\ref{eq:Wxl2}), we use the approximant $\widetilde{W}^{[2|1]}$ and its series expansion, for small $x$, is given by:
	\begin{align}\label{eq:W21}
	\widetilde{W}^{[2|1]} &= \frac{1 + A_1x + A_2 x^2}{1 - B_1 x} \, ,\nn\\
	&=1+x (A_1+B_1)+x^2 \left(A_1 B_1+A_2+B_1^2\right)\,\nn\\&+x^3 \left( A_1 B_1^2+A_2 B_1+B_1^3\right)+x^4 (A_1 B_1^3+A_2 B_1^2\,\nn\\&+B_1^4)+O\left(x^5\right)\, .
	\end{align}
	~Now, first we solve for $A_1$, $A_2$ and $B_1$ in terms of known $R_1,R_2$ and $R_3$ of eq(\ref{eq:Wxl2}) and the solutions are:
	\begin{align}
	A_1&= \frac{R_1 R_2-R_3}{R_2}\, ,\\
	A_2&= \frac{R_2^2-R_1 R_3}{R_2}\, ,\\
	B_1&= \frac{R_3}{R_2}\, .
	\end{align}
In the next step, the coefficient of $x^4$ in eq(\ref{eq:W21}) is taken as the prediction for $R_4$ which, in terms of the lower order $R_i$'s, can be written as:
	\begin{equation}
	R_4^{\text{pred}}=\frac{R_3^2}{R_2}\, .
		\end{equation}
In large $L$ limit, we get the the following form for $R_4^{\text{pred}}$ compatible with eq(\ref{eq:Rs}) as:
	\begin{equation}
	R_4^{\text{pred}} = T_{4,0,0}^{\text{Pad\'e}} + T_{4,1,0}^{\text{Pad\'e}} L + T_{4,2,0}^{\text{Pad\'e}} L^2 + T_{4,3,0}^{\text{Pad\'e}} L^3 + T_{4,4,0}^{\text{Pad\'e}} L^4 \, .
	\end{equation}
These predictions can be further improved by knowing the asymptotic behaviour of the series, and this extra piece of information is taken as input to get a more precise approximation. This improvement is termed as APAP and can be found in ref~\cite{Ellis:1996zn}.

The error associated with such approximation in asymptotic limit~\cite{Ellis:1996zn,Chishtie:2001mf} is given by:
	\begin{equation}
	\delta_{[N/M]}\approx - \frac{N^M A^{M}}{D^{M}}\, ,
	\end{equation}
where $D= N+M(1+a_p)+b_p$ and $A, a_p, b_p$ are fitting parameters. We get APAP results in terms of $R_4^{\text{pred}}$ by:
	\begin{equation}
	R_4^{\rm APAP} =\frac{R_4^{\text{pred}}}{(1+\delta_{[N/M]})}\, .
	\end{equation}
Repeating the procedure for known lower order $R_i's$, we can fix the constants $A, a_p$ and $b_p$ for a fixed $M$.  It is worth to mention that among different choice of Pad\'e approximant for APAP, for a given order, $M=1$ and $a_p=0,~b_p=0$ gives best result compatible with RG for the static energy and hence this particular choice is used in this article.

Following the procedure explained above, we get the four-loop Pad\'e prediction in the large $L$ limit  as:
	\begin{align}\label{ps0}
	T_{4,0,0}^{\text{Pad\'e}}\nonumber & = \frac{13 \beta_1 T_{1,0,0}^3}{\beta_0}+\frac{169}{9}\frac{ \beta_1^2 T_{1,0,0}^2}{ \beta_0^2}-\frac{62 }{9 }\frac{\beta_1 \beta_2 T_{1,0,0}}{\beta_0^2} \nonumber \\ &+\frac{43}{9}\frac{\beta_1 \beta_2}{\beta_0}+\frac{313}{27}\frac{\beta_1^3 T_{1,0,0}}{ \beta_0^3}-\frac{160}{9}\frac{ \beta_1 T_{2,0,0} T_{1,0,0}}{ \beta_0}\nonumber \\ &-\frac{62}{9}\frac{ \beta_2 T_{1,0,0}^{2}}{ \beta_0 }+\frac{43}{9}\frac{\beta_1 T_{3,0,0}}{\beta_0} + \frac{62}{9}\frac{ \beta_2 T_{2,0,0}}{ \beta_0}-\frac{55}{9}\frac{ \beta_1^2 \beta_2}{ \beta_0^3}\nonumber \\ & - \frac{571}{36}\frac{ \beta_1^2 T_{2,0,0}}{ \beta_0^2}+\frac{52}{9} T_{1,0,0}^4 -\frac{113 }{9}T_{2,0,0} T_{1,0,0}^2 \nonumber \\ & +\frac{46 }{9} T_{2,0,0}^2+ \frac{1429}{324}\frac{\beta_1^4}{\beta_0^4}+\frac{19}{9}\frac{\beta_2^2}{\beta_0^2}+\frac{8 }{3} T_{3,0,0} T_{1,0,0}\, ,
	\end{align}
	\begin{align}
	T_{4,1,0}^{\text{Pad\'e}} &= \frac{8}{3} \beta_0 T_{3,0,0}+\frac{61}{9} \beta_1 T_{2,0,0}-2 \beta_0 T_{1,0,0}^3+\frac{8}{3} \beta_2 T_{1,0,0}\nonumber \\ &-\frac{461}{108 }\frac{ \beta_1^3}{\beta_0^2} -\frac{34}{9} \beta_1 T_{1,0,0}^2 +\frac{10}{3} \beta_0 T_{2,0,0} T_{1,0,0}+\frac{43}{9}\frac{\beta_1 \beta_2}{\beta_0}\nonumber \\ &-\frac{43}{18}\frac{\beta_1^2 T_{1,0,0}}{ \beta_0}\label{ps1}\, ,
	\end{align}
	\begin{align}
	T_{4,2,0}^{\text{Pad\'e}} &= \frac{17}{3} \beta_0^2 T_{2,0,0}+\frac{22}{3} \beta_1 \beta_0 T_{1,0,0}+\frac{8}{3} \beta_2 \beta_0+\frac{79}{36} \beta_1^2\,\nn\\ &+
	\frac{1}{3} \beta_0^2 T_{1,0,0}^2 \label{ps2}\, ,
	\end{align}
	\begin{align}\label{ps3}
	T_{4,3,0}^{\text{Pad\'e}}= 4 T_{1,0,0} \beta_0^3+\frac{13}{3} \beta_0^2 \beta_1, \quad T_{4,4,0}^{\text{Pad\'e}} =\beta_0^4\, .
	\end{align}

We can see that for $T_{4,4,0}$ and $T_{4,4,3}$ the predictions from Pad\'e and the renormalization group are in perfect agreement. For the other RG-accessible coefficient $T_{4,1,0}$ and $T_{4,2,0}$, the predictions are different. However, numerical difference for $T_{4,2,0}$ is not more than 2.2\% for active quark flavours $n_f\le 6$. However, $T_{4,1,0}$ has larger deviations $\ge2\%$ for $n_f>2$ from RGE prediction. For this reason, we will restrict our discussion in next sections only to two active flavours for the four-loop. The Pad\'e prediction for the unknown constant term at the four-loop order for the static potential can be obtained by setting $T_{i,j,k}\xrightarrow{k>0}0$ and $\delta T^{\rm us}_{i,0,0}\rightarrow 0$. \par	
Interestingly in the large $n_f$ limits both the Pad\'e approximant and solutions of RGE for RG-accessible coefficients $T_{4,i,0}$ give the same values
	\begin{align}
	T_{4,1,0}^{\text{Pad\'e}} &\xrightarrow{n_f\rightarrow\infty}\frac{125}{8748}n_f ^4\, , \quad &T_{4,1,0}^{\text{RG}}&\xrightarrow{n_f\rightarrow\infty}\frac{125}{8748}n_f^4\, ,\nonumber\\
	T_{4,2,0}^{\text{Pad\'e}}&\xrightarrow{n_f\rightarrow\infty}\frac{25 }{1944}n_f^4\, , \quad &T_{4,2,0}^{\text{RG}}&\xrightarrow{n_f\rightarrow\infty}\frac{25}{1944}n_f^4\, ,\nonumber\\
	T_{4,3,0}^{\text{Pad\'e}}&\xrightarrow{n_f\rightarrow\infty}\frac{5}{972} n_f^4\, , \quad &T_{4,3,0}^{\text{RG}}&\xrightarrow{n_f\rightarrow\infty}\frac{5}{972}n_f^4\, ,\nonumber\\
	T_{4,4,0}^{\text{Pad\'e}}&\xrightarrow{n_f\rightarrow\infty}\frac{1}{1296}n_f^4\, ,\quad &T_{4,4,0}^{\text{RG}}&\xrightarrow{n_f\rightarrow\infty}\frac{1}{1296}n_f^4\nonumber\, .
	\end{align}
For RG-inaccessible coefficient $$T_{4,0,0}^{\text{Pad\'e}}\xrightarrow{n_f\rightarrow\infty}\left(\frac{-5 T_f n_f}{9}\right)^4.$$
In fact, similar pattern has been observed for known lower orders: $$T_{i,0,0}^{\text{Pad\'e}}\xrightarrow{n_f\rightarrow\infty}\left(\frac{-5 T_f n_f}{9}\right)^i.$$ ~Now we have an estimate for $T_{4,0,0}$ but the truncated perturbation series also suffers from scale dependence. The scale sensitivity of the perturbative series can be minimized using RG-summation of running logarithms and the procedure is discussed in the next section.

\section{RG Improvement in the momentum space\label{sec:RG_sum}}
The issue with the perturbative series in the QCD is to account for the RG running of all the parameters. The optimal renormalization method advocated in refs~\cite{Ahmady:2002fd,Ahmady:2002pa} accounts for the RG running by summation of all the RG-accessible logarithms. The RG-accessible logarithms at each order in the perturbation theory are defined as the leading and the next-to-leading logarithms that can be accessed through the processes-dependent the RGE. Resummation becomes interesting from the three-loop order due to presence of the ultrasoft terms and this issue is discussed for static energy for the first time in this article.

The perturbative series in question is 
\begin{align}
	&W(x,L)=\sum _{i=0}^n \sum _{j=0}^i \sum _{k=0}^{\tiny{\substack{(i-j-2)\\ \times \theta (i-j-3)}}} x^{i+1} L^j \log ^k(x) T_{i,j,k}
	\label{eq:eq1T2}\, ,
\end{align}
where the series coefficients are $T_{i,j,k}$. To obtain RG-summed perturbation series which we call $W^{(n)}_{\rm RG\Sigma}$, we rewrite $W(x,L)$ as
\begin{align}
	&W^{(n)}_{\rm RG\Sigma} = \sum _{i=0}^n \sum _{k=0}^{(i-2) \theta (i-3)} x^{i+1} \log ^k(x) S_{i,k}(x L)\, ,
	\label{eq:RGsummed}
\end{align}
where intermediate quantities $S_{i,k}(x L)$ are the resummed series obtained by summing terms:
\begin{equation}\label{eq:S}
\sum _{n=i}^{\infty} (x~L )^{n-i} T_{n,n-i,k}\, .
\end{equation}
We substitute eq(\ref{eq:eq1T2}) in eq(\ref{RG_W}) which leads to a recursion relation between the series coefficients. We multiply the recursion relation with $(x L)^{k-1}$ with appropriate $k$ and sum it from $n=k$ to infinity, which following eq(\ref{eq:S}), give differential equations for $S_{i,k}(x L)$. The solution to these differential equation results into the closed form expression for $S_{i,k}(x L)$.\par
The RG-summed solutions $S_{i,0}(x L)$ are calculated to the two-loop order in ref~\cite{Ahmady:2002fd} and are given below:
		\begin{align}
		S_0=&\frac{1}{w}\,,\quad S_1=w^{-2}\left(T_{1,0,0}-\tilde{B}_1 L_w\right)\,,\label{eq:S0}\\ 
		S_2=&w^{-2}\left(\tilde{B}_1^2-\tilde{B}_2\right)+w^{-3}\bigg[T_{2,0,0}-\tilde{B}_1^2+\tilde{B}_2\nn\\&-\tilde{B}_1 L_w \left(\tilde{B}_1+2 T_{1,0,0}\right)+\tilde{B}_1^2 L_w^2\bigg]\,,\label{eq:S2}
		\end{align} 
		where $w=(1-\beta_{0} u)$ and $\tilde{B}_i=\beta_i/\beta_0$ and $L_w\equiv\log(w)$. The RG-summation for $e^+ e^-$ process to three-loop order is also discussed in ref~\cite{Ahmady:2002pa} which is a special case of the results of this article in the limit where ultrasoft coefficients are taken zero. \par
The static energy requires a new series representation, compatible with the RGE, to incorporate the ultrasoft logarithms. These logarithms form a separate recurrence relation among the coefficients. The RG-summation of ultrasoft terms at the three-loop order is obtained by collecting the coefficients of $x^n L^{n-4}\log(x)$ in eq(\ref{eq:rg_W}) which results in the following recurrence relation among the coefficients:
	\begin{align}
	(n-3) T_{n,n-3,1}-n \beta_0 T_{n-1,n-3,1}=0\, .
	\end{align}
	Collecting $x^n L^{n-4}$ terms, we get the following recurrence relation:
	\begin{align}
	&(n-3)T_{n,n-3,0}-n\beta_{0} T_{n-1,n-4,0} -\beta_0T_{n-1,n-4,1}\nonumber\\&-(n-1)\beta_1T_{n-2,n-4,0} -(n-2)\beta_2 T_{n-3,n-4,0}\nonumber\\&-(n-3) \beta_3T_{n-4,n-4,0}=0\, .
	\end{align}
Notice that presence of the $\beta_0T_{n-1,n-4,1}$ terms in the above equation is new and differs from $e^+ e^-$ case in ref~\cite{Ahmady:2002pa}.

For the four-loop order, coefficients of $x^n L^{n-5}\log^2(x)$ terms give the following recurrence relation for ultrasoft terms:
	\begin{align}
	(n-4)T_{n,n-4,2}-n \beta _0 T_{n-1,n-5,2}=0\, .
	\end{align}
Collecting $x^n L^{n-5}\log(x)$ terms, we get the following recurrence relation:
	\begin{align}
	(n-4)&T_{n,n-4,1}-(n-1) \beta _1 T_{n-2,n-5,1}\nonumber\\&-n \beta _0 T_{n-1,n-5,1}-2 \beta _0 T_{n-1,n-5,2}=0\, .
	\end{align}
Collecting $x^n L^{n-5}$ terms we get the following recurrence relation:
	\begin{align}
	&(n-4)T_{n,n-4,0}-(n-4)\beta _4 T_{n-5,n-5,0}-(n-3)\nonumber\\&\times \beta _3 T_{n-4,n-5,0}-(n-2) \beta _2 T_{n-3,n-5,0}-(n-1) \nonumber\\&\times \beta _1 T_{n-2,n-5,0}-\beta _1 T_{n-2,n-5,1}-n \beta _0 T_{n-1,n-5,0}\nonumber\\&-\beta _0 T_{n-1,n-5,1}=0\, ,
	\end{align}
multiplying $u^{k-1}$, where $k=n-3$ for three-loop and $k=n-4$ for the four-loop, to the recurrence relations and then summing $n$ from $k$ to $\infty$ gives separate differential equations for the above recurrence relations. All these recursion relations can be written in a general-differential equation for the static energy as:
	\begin{align}
	&\sum _{i=0}^n \Big(\theta (i-k,-k)-\theta (i-k-2,k-1)\Big)\nonumber\\&\times \Bigg( (\delta _{i,n}-u \beta (n-i))\frac{d}{d u}S_{i,k}(u) -(i+1) \beta (n-i) S_{i,k}(u)\nonumber\\&-(k+1) \theta (i-k-3) \beta (n-i) S_{i,k+1}(u)\Bigg)=0\, .
	\end{align}
Solution to different RG-summed series to the order we are interested in are given by:
	\begin{align}
	&S_{3,1}(w)=\frac{T_{3,0,1}}{w^4}, \quad S_{4,2}(w)=\frac{T_{4,0,2}}{w^5}\, ,\label{eq:S42}\\
	S_{4,1}(w)&=\frac{1}{w^5}\left(-4 \tilde{B}_1 T_{3,0,1} L_w-2 T_{4,0,2} L_w+T_{4,0,1}\right)\,.
	\end{align}
Similarly, other solutions are:
		\begin{align}
		S_{3,0}(w)&=\frac{1}{w^4}\left(T_{3,0,0}-2 \tilde{B}_1^2 T_{1,0,0}+2 \tilde{B}_2 T_{1,0,0}-\frac{\tilde{B}_1^3}{2}+\frac{\tilde{B}_3}{2}\right)\nonumber\\&+\frac{L_w}{w^4} \left(-2 \tilde{B}_1^2 T_{1,0,0}-3 \tilde{B}_1 T_{2,0,0}+2 \tilde{B}_1^3-3 \tilde{B}_2 \tilde{B}_1-T_{3,0,1}\right)\nonumber\\&+\frac{L_w^2}{w^4} \left(3 \tilde{B}_1^2 T_{1,0,0}+\frac{5 \tilde{B}_1^3}{2}\right)-\frac{\tilde{B}_1^3 L_w^3}{w^4} +\frac{1}{w^3}\bigg(2 \tilde{B}_1^2 T_{1,0,0}\nonumber\\&-2 \tilde{B}_2 T_{1,0,0}+\tilde{B}_1^3-\tilde{B}_2 \tilde{B}_1\bigg)+\frac{L_w}{w^3}\left(2 \tilde{B}_1 \tilde{B}_2-2 \tilde{B}_1^3\right) \nonumber\\& +\frac{1}{w^2}\left(-\frac{\tilde{B}_1^3}{2}+\tilde{B}_2 \tilde{B}_1-\frac{\tilde{B}_3}{2}\right)\, ,
		\end{align}
\begin{widetext}
\begin{align}
		S_{4,0}(w)&=\frac{1}{w^5}\left(T_{4,0,0}-\tilde{B}_1^3 T_{1,0,0}-3 \tilde{B}_1^2 T_{2,0,0}+\tilde{B}_3 T_{1,0,0}+3 \tilde{B}_2 T_{2,0,0}+\frac{7 \tilde{B}_1^4}{6}-3 \tilde{B}_2 \tilde{B}_1^2-\frac{1}{6} \tilde{B}_3 \tilde{B}_1+\frac{5 \tilde{B}_2^2}{3}+\frac{\tilde{B}_4}{3}\right)\nonumber\\&+\frac{L_w}{w^5}\left(6 \tilde{B}_1^3 T_{1,0,0}-3 \tilde{B}_1^2 T_{2,0,0}-8 \tilde{B}_2 \tilde{B}_1 T_{1,0,0}-4 \tilde{B}_1 T_{3,0,0}-\tilde{B}_1 T_{3,0,1}+4 \tilde{B}_1^4-3 \tilde{B}_2 \tilde{B}_1^2-2 \tilde{B}_3 \tilde{B}_1-T_{4,0,1}\right)\nonumber\\&+ \frac{L_w^2}{w^5}\left(7 \tilde{B}_1^3 T_{1,0,0}+6 \tilde{B}_1^2 T_{2,0,0}+4 \tilde{B}_1 T_{3,0,1}-\frac{1}{2} 3 \tilde{B}_1^4+6 \tilde{B}_2 \tilde{B}_1^2+T_{4,0,2}\right)+\frac{L_w^3}{w^5}\left(-4 \tilde{B}_1^3 T_{1,0,0}-\frac{1}{3} 13 \tilde{B}_1^4\right)+\frac{\tilde{B}_1^4 L_w^4}{w^5}\nonumber\\&+\frac{1}{w^4}\left(2 \tilde{B}_1^3 T_{1,0,0}+3 \tilde{B}_1^2 T_{2,0,0}-2 \tilde{B}_2 \tilde{B}_1 T_{1,0,0}-3 \tilde{B}_2 T_{2,0,0}-2 \tilde{B}_1^4+5 \tilde{B}_2 \tilde{B}_1^2-3 \tilde{B}_2^2\right)+\frac{L_w^2}{w^4}\left(3 \tilde{B}_1^4-3 \tilde{B}_1^2 \tilde{B}_2\right)\nonumber\\&+\frac{L_w}{w^4}\left(-6 \tilde{B}_1^3 T_{1,0,0}+6 \tilde{B}_2 \tilde{B}_1 T_{1,0,0}-5 \tilde{B}_1^4+5 \tilde{B}_2 \tilde{B}_1^2\right)+\frac{1}{w^2}\left(\frac{\tilde{B}_1^4}{3}-\tilde{B}_2 \tilde{B}_1^2+\frac{2}{3} \tilde{B}_3 \tilde{B}_1+\frac{\tilde{B}_2^2}{3}-\frac{\tilde{B}_4}{3}\right)\nonumber\\&+\frac{1}{w^3}\left(-\tilde{B}_1^3 T_{1,0,0}+2 \tilde{B}_2 \tilde{B}_1 T_{1,0,0}-\tilde{B}_3 T_{1,0,0}+\frac{\tilde{B}_1^4}{2}-\tilde{B}_2 \tilde{B}_1^2-\frac{1}{2} \tilde{B}_3 \tilde{B}_1+\tilde{B}_2^2\right)+\frac{L_w}{w^3}\left(\tilde{B}_1^4-2 \tilde{B}_2 \tilde{B}_1^2+\tilde{B}_3 \tilde{B}_1\right)\, .\label{eq:S40}
		\end{align}
	\end{widetext}

\begin{figure}[ht]
	\centering
	\includegraphics[scale=0.45]{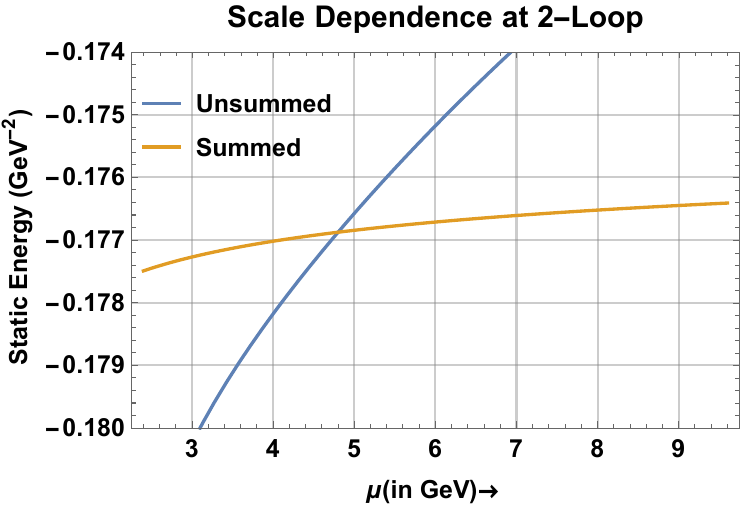}
	\includegraphics[scale=0.45]{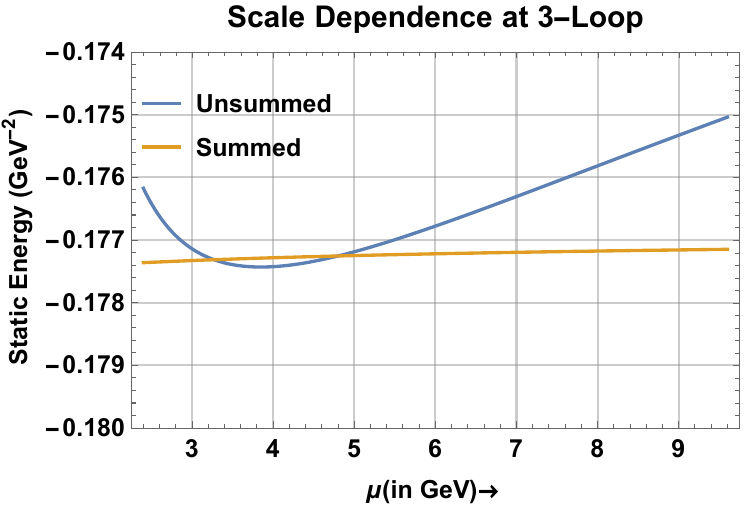}
	\includegraphics[scale=0.45]{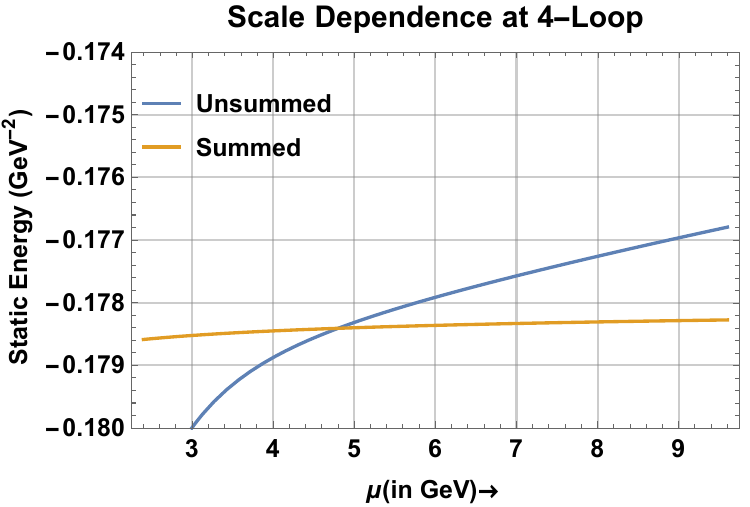}
	\caption{Renormalization scale dependence of the resummed and the unsummed static energy at different loops.}
	\label{fig:234_summed}
\end{figure}

The importance of resummation of all accessible logarithms can be seen in the figure(\ref{fig:234_summed}). The scale dependence of the RG-summed series eq(\ref{eq:RGsummed}) around momentum $p =\space m^{\overline{\rm MS}}_b=4.17\GeV$ is almost negligible in the range $m^{\overline{\rm MS}}_b/2 \le \mu \le 2 m^{\overline{\rm MS}}_b $ while the unsummed series in eq(\ref{E0exp}) has significant $\mu$ dependence. Scale sensitivity of unsummed series decreases order by order but the advantage of RG-summed series provides results less sensitive to scale with the same available information. This theoretical improvement provide us an opportunity to extract various parameters from available experimental data with less scale sensitivity.

\section{Restricted Fourier transform \label{sec:RFT}}	
Let us begin by noting that the ultrasoft part of the static energy is calculated in position space, whereas the perturbative part is carried out in momentum space. To make any phenomenological study, it is necessary to bring all the contributions of the static energy into the same space. Bringing the momentum space results in position space via a Fourier transform may seem natural, but such a transformation induces pathological contributions to the static energy. These undesirable contributions originate from the non-perturbative, small momentum modes, which must be removed explicitly. After removing such contributions, the convergence behaviour of the static energy improves drastically. This scheme was first addressed in ref~\cite{Karbstein:2013zxa} and was termed as the Restricted Fourier transform (RFT). This scheme has been already discussed in detail for the static energy to the three-loop. Here we discuss it very briefly and provide corresponding results for the four-loop order.

Note that the ultrasoft terms are already known to the four-loop order. Therefore, what we are providing is the complete calculation of the static energy in the position and momentum space to the four-loop order. The final result for the uncontrolled contribution to the position space static energy in RFT scheme is given in appendix~\ref{app:resVr} but an overview of the calculation is given here.

The position space version of the static potential, $V(r,\mu,\mu_{us})$, from momentum space potential, say $\tilde{V}(p,\mu,\mu_{us})$ to distinguish between the two basis, in RFT scheme is defined by:
	\begin{align*} 
	V\left(r,\mu,\mu_{us};\mu_{f}\right) &=\int_{\left[\mathbf{p} |>\mu_{f}\right.} \frac{\mathrm{d}^{3} p}{(2 \pi)^{3}} \mathrm{e}^{i \mathbf{p} \cdot \mathbf{r}} \tilde{V}(p,\mu,\mu_{us})\, , \\ &=V(r,\mu,\mu_{us})-\delta V\left(r, \mu_{us},\mu_{f}\right)\, ,
	\end{align*}
where $\mu_{f}$ is a perturbative scale chosen such that $\mu>\mu_f\gg \Lambda^{\overline{\textrm{MS}}}_{\textrm{QCD}}$ and uncontrolled terms, $\delta V\left(r,\mu,\mu_{us};\mu_{f}\right)$, in the potential is
	\begin{equation}
	\delta V\left(r, \mu,\mu_{us};\mu_{f}\right)=\int_{\left[\mathbf{p} |<\mu_{f}\right.} \frac{\mathrm{d}^{3} p}{(2 \pi)^{3}} \mathrm{e}^{i \mathbf{p} \cdot \mathbf{r}} \tilde{V}(p,\mu,\mu_{us};\mu_f)\, .
	\end{equation}
The static energy, given by eq(\ref{E0}), in RFT scheme is:
	\begin{equation}
	E_0\left(r, \mu;\mu_{f}\right)=V_s\left(r,\mu, \mu_{us};\mu_{f}\right)+\delta^{\rm us}\left(r,\mu, \mu_{us};\mu_{f}\right)\,.
	\end{equation}
	 The ultrasoft gluonic contribution to the static energy, at order $r^2$ in multipole expansion, is given by: 
	\begin{align}
	&\delta^{\rm us}(r,\mu,\mu_{us})=-i\frac{g^{2}}{N_{c}} T_{F} V_{A}^{2} \frac{r^{2}}{d-1}\,\nonumber\\ \times& \int_{0}^{\infty} d t \text{ } e^{-i t\left(V_{o}-V_{s}\right)} \times\left\langle 0\left|\mathbf{E}^{a}(t) \phi(t, 0)_{a b}^{\operatorname{ad} j} \mathbf{E}^{b}(0)\right| 0\right\rangle\, ,
	\end{align}
where $T_F=1/2$, $N_c$ is number of colors, $\phi(t, 0)_{a b}^{\operatorname{ad} j} $ is Wilson line in the adjoint representation connecting two points at temporal separation $t$, $ \mathbf{E}^{a/b}$ is choromoelectric field strength, and $V_A$ is matching coefficient which appear at order $r^2$ in multipole expansion. This quantity has been calculated to NLO in refs~\cite{Brambilla:1999qa,Brambilla:2006wp} and final result, sub-leading in $r$, is given by:
	\begin{align}
	\delta^{\mathrm{\rm us}}\left(r,\mu_{us}\right)=&-C_{F} \frac{\alpha_{s}(\mu_{us})}{\pi} \frac{r^{2}}{3} V_{A}^{2}(V_{o}(r)-V_{s}(r))^{3}\,\nonumber\\&\times\left(\delta^{\rm us}_{3-loop}+ \frac{\alphas(\mu_{us})}{\pi}\delta^{\rm us}_{4-loop}\right)\, ,
	\label{Vus4}
	\end{align}
	where 
	\begin{align}
	&\delta^{\rm us}_{3-loop}= 2 \log \left(\frac{V_{o}(r)-V_{s}(r)}{\mu_{\mathrm{us}}}\right)-\frac{5}{3}+2 \log 2\, ,
\shortintertext{and}
	&\delta^{\rm us}_{4-loop}=  C_{1} \log ^{2} \left(\frac{V_{o}(r)-V_{s}(r)}{\mu_{us}}\right)\,\nonumber\\&\hspace{2cm}+C_{2} \log \left(\frac{V_{o}(r)-V_{s}(r)}{\mu_{us}}\right)+D\, .
	\label{Vus2}
	\end{align}
Coefficients $C_1$, $C_2$ and $D$ are given in ref~\cite{Brambilla:2006wp} and we have presented them in appendix~\ref{app: loop_coef}.
	Next to the leading order expression for $V_{o}(r)-V_{s}(r)$ is given by:
	\begin{align}
	\left(V_{o}-V_{s}\right)&(r) =\frac{C_A \alpha _s(\mu )}{2 r}\newline\nn\\ &\times\bigg(1+\frac{\alpha _s(\mu )}{\pi } \left(T_{1,0,0}+(2 \gamma_E + \log(\mu ^2 r^2))\beta_0\right)\bigg)\nonumber \\ &+\order{\alpha_s^3(\mu)}\, ,
	\label{dVNLO}
	\end{align}
 and we can see that if we substitute eq(\ref{dVNLO}) to eq(\ref{Vus4}) then ultrasoft contributions to the three- and four-loop order can be obtained from NLO results.
	\par Due to the presence of the extra scale in the problem $\mu_{\mathrm{us}}$, it is desirable to expand all the quantities in terms of $\alpha_{s}(\mu)$. This expansion induces mixed logarithms at the four-loop of form $\frac{1}{r}\log(\mu^2_{us} r^2)\log(\mu^2 r^2)$ in the ultrasoft contributions. To simplify the calculation we can write:
		\begin{align}
	&\alpha_s(\mu_{us})=\alpha_s(\mu)\left(1+\frac{\alpha_s(\mu)}{\pi} \beta_0 \log \left(\frac{\mu^2}{\mu^2_{us}}\right)\right)+\order{\alpha_s^3(\mu)}\, ,\nonumber\\
	&\text{and}\nonumber\\
	&\log \left(r^2 \mu _{us}^2\right)=\log \left(\mu ^2 r^2\right)-\log \left(\frac{\mu ^2}{\mu _{us}^2}\right)\, ,
	\end{align}
	so that the ultrasoft scale will appear only in running logarithms.
	\par 
The uncontrolled term for $\left(V_{o}-V_{s}\right)(r)$ in RFT scheme, in next to the leading order is given by:
	\begin{align}
	\delta\left(V_{o}-V_{s}\right)&(r,\mu_f)=C_A \mu_f x \Bigg[H_1+x \Bigg(H_1 T_{1,0,0}\nonumber\\&+\beta _0 \big(2 H_2+H_1 L_{\mu_f}\big)\Bigg)\Bigg]+\order{x^3}, 
	\end{align}
where $L_{\mu_f}=\log(\frac{\mu^2}{\mu^2_f})$ and $H_i's$ functions  defined by eq(\ref{hyp}) and are proportional to generalized hypergeometric function. The only matching coefficient left is $V_A$ and it is taken as $$V_A=\frac{C_A \alpha_{s}}{2 r}\frac{1}{V_o-V_s}+\order{\alpha_{s}},$$ same as calculated in ref~\cite{Karbstein:2013zxa}. These contributions will enter in eq(\ref{Vus4}) and the final expression is given in appendix~\ref{app:resVr}.

Pad\'e estimates and the ultrasoft contributions give us numerical estimate for $T_{4,0,0}=T_{4,0,0}^{\text{Pad\'e}}+\delta T_{4,0,0}^{\text{\rm us}}$. Now, the restricted version for the static potential, the ultrasoft contributions and the static energy in position space can be constructed using eq(\ref{FFT}) and eq(\ref{UnFT}).\par
\begin{figure}[ht]
		\includegraphics[scale=.8]{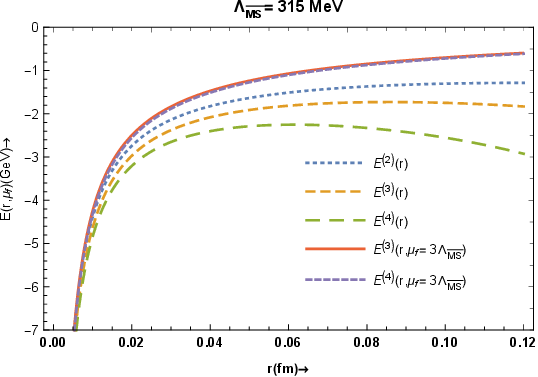}
		\caption{Restricted and unrestricted static energy to a given order(in superscript). $\Lambda^{\overline{\textrm{MS}}}_{\textrm{QCD}}=315\MeV$ is used as input.}
		\label{fig:restE}
	\end{figure}

 
Using $\Lambda^{\overline{\textrm{MS}}}_{\textrm{QCD}}=315 \MeV$, we can see in figure(\ref{fig:restE}) that the four-loop corrections to the static energy in RFT scheme makes a very small contribution but, without removing pathological contributions, behaviour of the same quantity in unrestricted scheme has very bad for $r>0.05~{\rm fm}$ at the four-loop.
	\par
	In the next section, we have performed the extraction of $\Lambda^{\overline{\textrm{MS}}}_{\textrm{QCD}}$ in momentum space using LQCD inputs from ref~\cite{Karbstein:2018mzo}.
	
\section{Fitting Perturbative results to lattice data \label{sec:lat_input}}
In this section, we fit the perturbative static energy to the lattice data and extract the value of the $\Lambda^{\overline{\textrm{MS}}}_{\textrm{QCD}}$ from RG-summed and unsummed case. Due to absence of all order results, the extracted quantity $\Lambda^{\overline{\textrm{MS}}}_{\textrm{QCD}}$ depends on the choice of renormalization scale. To reduce the scale dependence, we exploit the optimal renormalized perturbative static energy to extract $\Lambda^{\overline{\textrm{MS}}}_{\textrm{QCD}}$.
The parametrization of lattice data to the Cornell potential is given in position space in ref~\cite{Karbstein:2018mzo} for two-flavour QCD 
	\begin{equation}
	E_{\rm lat}(r)=V_0-\frac{\alpha}{r}+\sigma r,
	\label{lp}
	\end{equation}
where $\alpha =0.326\pm0.005$ and $\sigma= 7.52\pm0.55 {\rm fm}^{-2}$ are constants for $n_f=2$ flavour. The parameter $V_0$ is a potential offset needed to match to the perturbative static energy. However, $V_0$ is not needed in case of momentum space analysis\cite{Karbstein:2018mzo}. It is important to note that these parameters are correlated and have $cor(\alpha,\sigma )=-0.17$ which has to be taken account in sampling from normal distribution.\par 
The Fourier transform of lattice static energy, in eq(\ref{lp}), to momentum space is given by:
	\begin{equation}
	E_{\rm lat}(p)=-\frac{4 \pi \alpha}{p^2}-\frac{8\pi \sigma}{p^4}\, .
	\label{lpm}
	\end{equation}
To extract $\Lambda^{\overline{\textrm{MS}}}_{\textrm{QCD}}$, we minimize the square-deviation~\cite{Karbstein:2018mzo}:
	\begin{equation}
	\Delta(\Lambda^{\overline{\textrm{MS}}}_{\textrm{QCD}})=	\int_{p_{min}}^{p_{max}}dp \left(E_{0}\left(p,\mu\right)- E_{\rm lat}\left(p\right) \right)^2 .
	\end{equation}
It should be noted that the strategy for sampling is analogous to the one discussed in ref~\cite{Karbstein:2018mzo}. The matching region is chosen $1500\MeV\le p\le3000 \MeV$ and momentum values are randomly sampled from a uniform distribution with $p_{min}\subset[1500,2250]\MeV$ and $p_{max}\subset[2250,3000]\MeV$ such that $p_{max}-p_{min}\ge375\MeV$. It is important to note that five-loop running of $\alpha_{s}$ is used to extract $\Lambda^{\overline{\textrm{MS}}}_{\textrm{QCD}}$.\par

\begin{figure}[ht]
		\includegraphics[scale=.8]{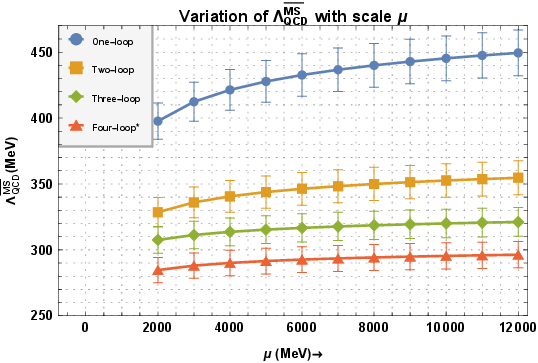}
		\caption{Renormalization scale dependence of $\Lambda^{\overline{\textrm{MS}}}_{\textrm{QCD}}$ at different loop order with error bars. The full Pad\'e estimated value for $T_{4,0,0}$ is used from eq(\ref{pade4}).}
		\label{sc_dep_lam}
	\end{figure}

Following the procedure explained above, the scale dependence of the extracted (for 500 samples for $\{p,\alpha,\sigma\}$) $\Lambda^{\overline{\textrm{MS}}}_{\textrm{QCD}}$ can be seen from figure(\ref{sc_dep_lam}) at different loop order and at different renormalization scales.
		 \par
In table~\ref{lambda_3}, we present our determinations of $\Lambda^{\overline{\textrm{MS}}}_{\textrm{QCD}}$ for different orders of the perturbative static energy and for different choices of the renormalization scale. In addition to reduction in the errors, the central values are closer to each other for different choices of $\mu$ at higher order in the perturbation series.
	\begin{table}[ht]
		\begin{center}
			\begin{tabular}{|c|c|c|c|c|}
				\hline
				\multirow{3}{*}{Loop} & \multicolumn{4}{c|}{$\Lambda^{\overline{\textrm{MS}}}_{\textrm{QCD}}(\text{in }\MeV)$} \\
				\cline{2-5}
				&\multicolumn{1}{c|}{unsummed}&\multicolumn{3}{c|}{RG-Summed}\\
				\cline{2-5}
				& $\mu=p$& $\mu=2.25\GeV$& $\mu=4.17\GeV$&$\mu=6.5 \GeV$\\
				\hline			
				1&$398.3\pm13.7$&$402.3\pm13.0$& $422.7\pm15.0$&$434.9\pm15.8$\\
				2&$328.8\pm11.2$&$330.9\pm11.0$&$341.3\pm11.6$&$347.4\pm12.0$\\
				3&$307.3\pm10.9$&$308.4\pm10.6$&$313.7\pm10.9$&$317.0\pm11.1$\\
				\hline
			\end{tabular}
		\end{center}
		\caption{\label{lambda_3}$\Lambda^{\overline{\textrm{MS}}}_{\textrm{QCD}}$ at different loop-orders}
	\end{table}\par
At the four-loop order (for 2000 samples of $\alpha,\sigma,p$), we make the following three choices for the unknown coefficient $T_{4,0,0}$: 
	\begin{equation}
	T_{4,0,0}=\{T_{4,0,0}^{\text{Pad{\' e}}}+\delta T_{4,0,0}^{us},\frac{T_{4,0,0}^{\text{Pad{\' e}}}}{2}+\delta T_{4,0,0}^{us},\frac{T_{4,0,0}^{\text{Pad{\' e}}}}{4}+\delta T_{4,0,0}^{us}\}\,,
	\label{pade4}
	\end{equation}
and the corresponding results are presented in Table~\ref{lambda_4}.
	\begin{table}[h]
		\begin{center}
			\begin{tabular}{|c|c|c|c|c|}
				\hline
				\multirow{3}{*}{$T_{4,0,0}$} & \multicolumn{4}{c|}{$\Lambda^{\overline{\textrm{MS}}}_{\textrm{QCD}}(\text{in }\MeV)$} \\
				\cline{2-5}
				&\multicolumn{1}{c|}{unsummed}&\multicolumn{3}{c|}{RG-Summed}\\
				\cline{2-5}
				& $\mu=p$& $\mu=2.25\GeV$&$\mu=4.17\GeV$&$\mu=6.5 \GeV$\\			
				\hline			
				I&$284.6\pm10.0$&$285.6\pm9.6$&$290.1\pm9.7$&$292.7\pm9.8$\\ 
				II&$296.3\pm10.3$&$297.0\pm10.0$&$300.0\pm10.1$&$302.0\pm10.2$\\
				III&$302.9\pm10.4$&$303.4\pm10.3$&$305.5\pm10.3$&$307.0\pm10.4$\\ 
				\hline
			\end{tabular}
		\end{center}
		\caption{\label{lambda_4}$\Lambda^{\overline{\textrm{MS}}}_{\textrm{QCD}}$ at different loop-order for choices of $T_{4,0,0}$ according to eq(\ref{pade4}).}
	\end{table}
	
The RG-summed and unsummed series give very similar values of $\Lambda^{\overline{\textrm{MS}}}_{\textrm{QCD}}$ in case of renormalization scale is chosen in the middle of the matching region. However, the RG-summed static energy provides better fit to the lattice static energy than the unsummed one which can be seen in Table~\ref{lambda5} below:
	\begin{table}[ht]
		\begin{center}
			\begin{tabular}{|c|c|c|c|}
				\hline
				\multirow{3}{*}{Loop} & \multicolumn{3}{c|}{Percent of cases (in \%) with $\left(\Delta_S(\mu)-\Delta_U\right) <0$} \\
				\cline{2-4}
				& $\mu=2.25\GeV$& $\mu=4.17\GeV$&$\mu=6.5 \GeV$\\
				\hline
				1&85.45& 81.60& 79.70\\
				2&97.10& 95.50& 94.75\\
				3&99.94&99.92& 99.84\\	
				\hline
			\end{tabular}
		\end{center} 
		\caption{\label{lambda5} Percentage of cases where the RG-summed series gives small square-deviation compared to the unsummed one at different loop order.}
	\end{table}
	
The RG-summed static energy gives better fit and this improvement persist even at next higher order. For example, at the four-loop order, we have found that the RG-summed static energy produces significantly better fit to lattice static energy compared to unsummed perturbative static energy for different cases when we choose $T_{4,0,0}$ according to eq(\ref{pade4}).
	\section{Discussions \label{sec:conc}}
The static energy between a heavy quark and anti-quark is an important quantity in QCD as it can be calculated in both perturbation theory and in LQCD simulations. Existence of a matching region, where both perturbation theory and lattice agree, provides an opportunity to extract the parameters of the theory. If we assume the light quarks are massless and the heavier one decouples then the only parameter left in the theory is strong coupling constant. Precise calculation of the static energy thus becomes very important goal for precision physics involving the strong interactions.

The perturbative static energy involves the computation of all the Feynman diagrams at that order and the computer codes were used in refs~\cite{Anzai:2009tm,Smirnov:2009fh} for the three-loop numerical calculation. At this order, virtual emission of the ultrasoft gluons, with energy and momentum smaller than $1/r$, capable for changing singlet state to octet state and vice versa also appear. Such emissions induce infrared divergence in the static potential.

The gluon fields are multipole expanded about inter-quark separation and hence their energy and momentum are restricted to ultrasoft scale. This scale acts as source of ultraviolet divergence for gluonic contributions. However the static energy, which is sum of perturbative potential and contribution of ultrasoft gluons, is divergence free quantity as two divergences get cancelled with each other. This cancellation induces non-analytic terms($\sim\alpha^n_{s}(1/r)\log^m\left(\alpha_{s}(1/r)\right)$) \cite{Appelquist:1977es} in the static energy. The ultrasoft terms\cite{Brambilla:1999qa} and their resummation is known to NLO \cite{Pineda:2000gza,Brambilla:2009bi} which contributes at the the three- and four-loop order but the perturbative singlet potential is known only to the three-loop\cite{Anzai:2009tm,Smirnov:2009fh,Lee:2016cgz}. 

\subsection{Pad\'e estimate}
In this article, we have extended the results for static energy to the four-loop order using the renormalization group and Pad\'e approximant. Among the seven coefficients at the four-loop, the estimates for $T_{4,3,0}$ and $T_{4,4,0}$, using APAP, are in exact agreement with solutions of renormalization group. Deviation for coefficient $T_{4,2,0}$ is within 2.2\% for $n_f\le6$ but $T_{4,1,0}$ deviates more than 2\% for $n_f>2$. Since, the deviation is less than 2\% for RG-accessible coefficient for light flavour, we have used APAP estimate for $T_{4,0,0}$ to extract the $\Lambda^{\overline{\textrm{MS}}}_{\textrm{QCD}}$ to the four-loop order with different choices given in eq(\ref{pade4}). The large-$n_f$ limits for RG-accessible terms from both methods are in perfect agreement. There are also some contribution to $T_{4,0,0}$ from ultrasoft terms which demands this quantity must be Fourier transformed in order to get complete the RG-inaccessible terms in momentum space. 

\subsection{Position space improvement}
The static energy from LQCD simulations are mostly parametrized in position space and hence the perturbative static energy is Fourier transformed to the position space. This quantity in position space suffers from pathological contributions stemming from the non-perturbative regions and has to be removed explicitly. This is achieved using the Restricted Fourier transform advocated in ref~\cite{Karbstein:2013zxa} which improves the convergence behaviour for $r\sim0.12$ fm. The ultrasoft and the static potential has explicit dependence on another scale $\mu_{\mathrm{\rm us}}$ which is absent in total energy. This scale-dependence should also be cancelled for the static energy in the RFT scheme. Final expression for uncontrolled contributions to the static energy is provided in appendix~\ref{app:resVr}. The four-loop contribution to static energy in RFT scheme has very little effect and can be seen in figure(\ref{fig:restE}). The static energy in RFT scheme provided in the section~\ref{sec:RFT} for the four-loop order and can be used in future studies.
\subsection{RG-improvement of the static energy and $\Lambda^{\overline{\textrm{MS}}}_{\textrm{QCD}}$ extraction}
The RG-summed static energy in momentum space is used in this article to extract $\Lambda^{\overline{\textrm{MS}}}_{\textrm{QCD}}$ by fitting the static energy from perturbative to the static energy from LQCD from ref~\cite{Karbstein:2018mzo} for two active flavour. Its value for the three-loop from the RG-summed static energy is found to be $308.4\pm10.6 (2.25\GeV) \MeV$, $313.7\pm10.9 (4.17\GeV))\MeV$ and $317.0\pm11.1(6.5\GeV)\MeV$ where quantity in the parenthesis is the renormalization scale. For the unsummed static energy, this parameter is found to be $307.3\pm10.9\MeV$. The RG-summed version of static energy has been observed to provide not only the better fit to the the lattice energy but also giving less standard deviation if the renormalization scale is chosen in the middle of matching region. Similar trend also persist for next order but we have used less sample size since the exact calculations are not available. Our finding of $\Lambda^{\overline{\textrm{MS}}}_{\textrm{QCD}}$ from RG-summed and unsummed series agrees within error bars to the findings of ref~\cite{Karbstein:2018mzo}.

\section{Summary \label{sec:summary}}
To summarize, the QCD static potential is known to three-loop order, and the ultrasoft terms which first appear at the three-loop order are known to four-loop order. In section~\ref{sec:pert_pot}, we describe the perturbative and the ultrasoft part of the static static energy. The main results of this paper are the following
\begin{itemize}
\item In section~\ref{sec:RG_ser}, using the RGE we determine the RG-accessible coefficients at four-loop order which is shown in eq(\ref{RG_sol}).
\item The constant term of the four-loop coefficient can not be determined using RGE. In  section~\ref{sec:Pade_est}, we use the Pad\'e approximant method to obtain this term and is given in equation eq(\ref{ps0}). 
\item In section \ref{sec:RG_sum}, we apply for the first time the method of optimal renormalization to QCD static energy beyond two-loop order to sum up the RG-accessible running logarithms to all order in perturbation theory. The RG-summed series is defined in equation eq(\ref{eq:RGsummed}) and the subsequent quantities are given in eq(\ref{eq:S0}-\ref{eq:S2}), eq(\ref{eq:S42}-\ref{eq:S40}). The RG-summed series ensures the expected reduction in sensitivity to the renormalization scale as shown in figure(\ref{fig:234_summed}).
\item In section \ref{sec:RFT}, we use the Restricted Fourier Transform scheme to improve the convergence behaviour of the static energy in the position space to four-loop order. 
\item Using the RG-summed series eq(\ref{eq:RGsummed}) (see definition in eq(\ref{E0exp})) and the lattice QCD parametrization eq(\ref{lpm}), we fit the the QCD scale $\Lambda^{\overline{\textrm{MS}}}_{\textrm{QCD}}$ to the lattice data. Our fit results at different loop orders can be found Table~\ref{lambda_3}-\ref{lambda_4} and scale dependence at these orders in figure~\ref{sc_dep_lam}.
\item The uncertainties associated with our extraction of the $\Lambda^{\overline{\textrm{MS}}}_{\textrm{QCD}}$ is discussed in section \ref{sec:conc}.
\end{itemize}\par
In summary, we have used a variety of techniques, theoretical and numerical, and have rendered the picture of the QCD static energy as a very useful tool to obtain a clear handle on $\Lambda^{\overline{\textrm{MS}}}_{\textrm{QCD}}$ which is one of the fundamental parameters of the QCD, there by confirming the results in a large number of other studies. We have also studied the consistency of the picture by invoking Pad\'e approximants as well as renormalization group summation in order to achieve these ends. We also provide improvement in position space using RFT- scheme. Our findings in this article provide better control over variation of renormalization scale for finite order results available for the static energy. It also discusses scale dependence of the extracted $\Lambda^{\overline{\textrm{MS}}}_{\textrm{QCD}}$ at different orders of perturbation theory for the first time in this article as an application of the method.

\section{Acknowledgments} 
BA thanks D. Wyler and the University of Zurich, Switzerland for hospitality when this work started. BA was partly supported by the Mysore Sales International Ltd. Chair of the Division of Physical and Mathematical Sciences, Indian Institute of Science. DD would like to thank the DST, Government of India for the INSPIRE Faculty Award (grant no IFA16-PH170). AK is support by a fellowship from the Ministry of Human Resources Development, Government of India. We thank R. Sarkar for collaboration at an early stage of this investigation. In addition, we are grateful to N. Brambilla, P. Hegde, F. Karbstein, P. Lamba, H. Takaura and A. Vairo for invaluable discussions. We also thank S. Dey and S. Banik for help with numerical simulations. We are particularly grateful to N. Brambilla for reading the manuscript and making numerous suggestions that have improved it greatly.
		
\appendix
\begin{widetext}	

\section{The QCD $\beta$-functions \label{app:QCDbeta}}
The QCD beta function is given by $$\beta (x) \equiv - \sum_{i=0}^{\infty} \beta_i x^{i+2}$$ and $\beta_i$ are the coefficients beta-function at $(i+1)$-loop. The $\beta_i$ coefficients for $n_f$ active quark flavors up to five-loop order are \cite{vanRitbergen:1997va,Gross:1973id,Caswell:1974gg, Jones:1974mm,Tarasov:1980au,Larin:1993tp,Czakon:2004bu,Baikov:2016tgj,Herzog:2017ohr} 
	\begin{align}
	&\beta_0 = \frac{11}{4}-\frac{1}{6}n_f\, \quad 	\beta_1= \frac{51}{8} - \frac{19}{24}n_f\,,\quad
	\beta_2 = \frac{2857}{128} - \frac{5033}{1152} n_f + \frac{325}{3456}n_f^2\,, \\
	&\beta_3 = \frac{149753}{1536} - \frac{1078361}{41472} n_f + \frac{50065}{41472} n_f^2 + \frac{1093}{186624} n_f^3 + \frac{891}{64} \zeta(3) - \frac{1627}{1728} n_f \zeta(3) + \frac{809}{2592} n_f^2 \zeta(3)\,,
	\end{align}

	\begin{align}
	\beta_{4} =&\frac{621885 \zeta (3)}{2048}-\frac{144045 \zeta (5)}{512}+\frac{8157455}{16384}-\frac{9801 \pi ^4}{20480} + n_{f}\big(-\frac{1202791 \zeta (3)}{20736}+\frac{1358995 \zeta (5)}{27648}+ \frac{6787 \pi ^4}{110592} -\frac{336460813}{1990656}\big)& \nn \\ &+n_{f}^{2}\left(\frac{698531 \zeta (3)}{82944}-\frac{5965 \zeta (5)}{1296}-\frac{5263 \pi ^4}{414720}+\frac{25960913}{1990656}\right)+n_{f}^{3}\left(-\frac{24361 \zeta (3)}{124416}+\frac{115 \zeta (5)}{2304}+\frac{809 \pi ^4}{1244160}-\frac{630559}{5971968}\right)& \nn \\ &+n_{f}^{4}\left(\frac{1205}{2985984}-\frac{19 \zeta (3)}{10368}\right)\, .
	\end{align}	
	
\section{Running of The Perturbative QCD Coupling Constant\label{app:alpha_run}}
The running of the strong coupling constant in terms of known $\beta$ functions and the strong coupling at renormalization scale $\mu$ \cite{Jezabek:1998wk}, is given by:
		\begin{align}
		&x(p)\equiv\frac{\alpha_s(p)}{\pi} = x \Bigg(1+ x \beta _0 L+x^2 \left(\beta _1 L+\beta _0^2 L^2\right)+x^3 \left(\beta _2 L+\frac{5}{2} \beta _1 \beta _0 L^2+\beta _0^3 L^3\right)+x^4 \big(\beta _3 L+\big(\frac{3 \beta _1^2}{2}+3 \beta _0 \beta _2\big) L^2\nonumber\\&+\frac{13}{3} \beta _1 \beta _0^2 L^3+\beta _0^4 L^4\big)+x^5\Big( \beta _4 L+\big(\frac{7 \beta _1 \beta _2}{2}+\frac{7 \beta _0 \beta _3}{2}\big) L^2+\big(6 \beta _2 \beta _0^2+\frac{35}{6} \beta _1^2 \beta _0\big) L^3+\frac{77}{12} \beta _1 \beta _0^3 L^4+\beta _0^5 L^5\Big) \Bigg)+ \order{x^6}
			\label{alphasmu}
		\end{align}
where $L=\log(\mu^2/p^2)$. \par 
The couplant used to extract the $\Lambda^{\overline{\textrm{MS}}}_{\textrm{QCD}}$ at scale $\mu$ is given by:
		\begin{align}
		x(\mu)=&\frac{ y }{b_1}\Bigg(1-\ell y+y^2 \left(\frac{b_2}{b_1^2}+\ell^2-\ell-1\right)-y^3 \left(-\left(2-\frac{3 b_2}{b_1^2}\right) \ell+\frac{1}{2} \left(1-\frac{b_3}{b_1^3}\right)+\ell^3-\frac{5 \ell^2}{2}\right)\nonumber\\&+y^4 \left(-\left(\frac{3}{2}-\frac{6 b_2}{b_1^2}\right) \ell^2+\left(-\frac{3 b_2}{b_1^2}-\frac{2 b_3}{b_1^3}+4\right) \ell+\frac{b_4}{3 b_1^4}-\frac{b_2 \left(3-\frac{5 b_2}{3 b_1^2}\right)}{b_1^2}-\frac{b_3}{6 b_1^3}+\ell^4-\frac{13 \ell^3}{3}+\frac{7}{6}\right)\Bigg)+\order{y^6}
		\label{as_lam}
		\end{align}
		
		where $b_i\equiv \frac{\beta_i}{\beta_0}$, $\ell\equiv\log(\log(\mu^2/(\Lambda^{\overline{\textrm{MS}}}_{\textrm{QCD}})^2))$ and $y\equiv\frac{b_1}{\beta_0 \log(\mu^2/(\Lambda^{\overline{\textrm{MS}}}_{\textrm{QCD}})^2)}$.

\section{The QCD-Static Coefficients at Different Loop Order \label{app: loop_coef}}
The known results for the static energy is presented here. The coefficients of the perturbative part $V^{\rm pert}$ at different loop orders are listed below:
		\begin{align}
		\textbf{The one-loop terms:}\quad&T_{1,0,0}=\frac{31}{12}-\frac{5 n_f}{18} \,, \quad T_{1,1,0}=\beta _0 \,,\\
	\textbf{The two-loop terms:}\quad &T_{2,0,0}=28.5468-4.14714 n_f+\frac{25 n_f^2}{324}\, ,\quad T_{2,1,0}=2 T_{1,0,0} \beta _0+\beta _1\, ,\quad T_{2,2,0}=\beta _0^2\, ,\\
	\textbf{The three-loop terms:}\quad &T_{3,0,0}=209.884-51.4048 n_f+2.90609 n_f^2 -0.0214335 n_f^3\, ,\quad T_{3,3,0}=\beta _0^3\,,\nn \\ &T_{3,1,0}=2 T_{1,0,0} \beta _1+3 T_{2,0,0} \beta _0+\beta _2\,,\quad T_{3,2,0}=3 T_{1,0,0} \beta _0^2+\frac{5 \beta _1 \beta _0}{2}\,.
		\end{align}
 The ultrasoft contribution to the three-loop RG-inaccessible coefficient is given by: 
		\begin{align}
		\delta \widetilde{T}^{us} _{3,0,0}=\frac{1}{72} \pi ^2 C_A^3 (6 \ell_1-5)\,,
		\end{align}
and contribution to the four-loop is given by:
		\begin{align}
		\delta \widetilde{T}^{us}_{4,0,0}=&\frac{C_A^4\pi^2}{2592}\Big(18 \pi ^2 \gamma_E +141 \gamma_E-6 L_2 \left(66 \ell_1+6 \pi ^2+47\right)+198 L_2^2-3 \left(47+6 \pi ^2\right) L_{\pi }+72 \pi ^2 \ell_1+894 \ell_1\nonumber\\&+432 \zeta (3)-81 \pi ^2-1241\Big)+\frac{C_A^3\pi^2}{1728}\Big( 432 \ell_1T_{1,0,0}-216 T_{1,0,0}+60 \pi ^2 \beta _0+60 \gamma \beta _0-536 \beta _0+\left(165-60 \beta _0\right) L_{\pi }\nonumber\nonumber\\&+6 L_2 \left(-20 \beta _0+132 \ell_1+55\right)-396 L_2^2-144 \beta _0 \ell_1^2+480 \beta _0 \ell_1-1320 \ell_1+66 \pi ^2-165 \gamma +1474\Big)
		\end{align}
		where $\ell_1= \log(C_A \pi )+\gamma_E $, $L_{\pi}=\log(\pi)$ and $L_{2}=\log(2)$. The constant terms appearing in eq(\ref{Vus2}) can be found in the ref~\cite{Brambilla:2006wp} and are given by:
		\begin{align}
		C_1=& \frac{2}{3} \beta_0\,,\quad\quad
		C_2= \frac{1}{54} \bigg(C_A \left(-12 \pi ^2-149+66 \log (2)\right)+4 n_f T_f (10-6 \log (2))\bigg)\, ,\nn\\
		D=& \frac{C_A}{9}\Bigg(\bigg(\frac{9 \pi ^2}{4}+\frac{1241}{36}+\frac{11 \log ^2(2)}{2}-\frac{\gamma_E}{2}\left(\pi ^2 +\frac{47 }{6}\right)-12 \zeta (3)- \left(\pi ^2+17\right) \log (2)+\frac{1}{2} \pi ^2 \log (\pi )\nonumber+\frac{47 }{12}\log (\pi )\bigg)\nonumber\\&+ n_f T_f \left(\frac{5}{6} \left(\gamma_E +\log\left(64/\pi \right)\right)-\frac{\pi ^2}{3}-\frac{67}{9}-2 \log ^2(2)\right) \Bigg)\, .
		\end{align}
		
\section{Position Space Potential \label{app:Vr}}
The unrestricted Fourier integrals of logarithms to position space is given by:
		\begin{align}
		\int \frac{d^3 \mathbf{p} }{(2 \pi)^3} &e^{-\text{i} \mathbf{p}.\mathbf{r}}\frac{ 4 \pi }{\mathbf{q}^2} \log^{m} \left(\frac{\mu^2}{\mathbf{q}^2}\right) = \frac{1}{r} \sum_{j=0}^m {m \choose j} \log^{j}\left( \mu^2 r^2\right) \partial_{\eta}^{m-j}y\left(\eta\right)\vert_{\eta=0}
		\label{FFT}
		\end{align} 
and the RFT of these logarithms are given by:
		\begin{align}
		&\int \frac{d^3 \mathbf{p}}{(2 \pi)^3} e^{-i \mathbf{p}\mathbf{r}}\frac{4\pi}{\mathbf{q}^2} \log^m\left(\frac{\mu^2}{\mathbf{q}^2}\right)
		\Theta\left(\mu_{f}-|\mathbf{p}|\right)= -\frac{\mu_{f}}{\pi} \sum_{j=0}^{m}{m \choose j} \log ^{j}\left(\frac{\mu^{2}}{\mu_{f}^2}\right)(-2)^{m-j}\left[\partial_{\eta}^{m-j}f(\eta,r \mu_f) \right]_{\eta=0}
		\label{UnFT}
		\end{align}
Here $y(\eta)$ and $f(\eta,\beta)$ are given by:
		\begin{align}
		&y(\eta) \equiv e^{\tiny{ \left(2 \gamma_{E} \eta+\sum_{l=2}^{\infty} \eta^{l} \frac{(2^{\mathrm{l}}-1-(-1)^{l} )\zeta(l)}{l}\right)}}=\frac{\Gamma (1-2 \eta )}{\Gamma (1-\eta ) \Gamma (\eta +1)}\,,\nn \\
		&f(\eta,\beta) \equiv \frac{\Gamma(\eta)-\Gamma\left(\eta,\mathrm{i} \beta\right)}{\left(\mathrm{i}\beta\right)^{1+\eta}}+\mathrm{c.c.}
		\end{align}
and c.c. stands for complex conjugate. Writing $L_{\gamma}=2\gamma_E -\log (\mu^2 r^2)$, the unrestricted Fourier transforms for static potential without the ultrasoft terms can be written as:
		\begin{align}
			V(r)=\sum_{i=0}^{4} \sum_{j=0}^{i}x^{i+1} V_j(r)+\order{x^6}
		\end{align}
		\begin{align}
		V_0(r)=\frac{1}{r} \,, \quad V_1(r)=\frac{L_{\gamma}}{r} \,, \quad V_2(r)=\frac{1 }{ r}(L_{\gamma}^2+\frac{\pi ^2}{3}) \,, \quad 
		V_3(r)=\frac{L_{\gamma}}{r}(L_{\gamma}+\pi ^2) +\frac{16}{r} \zeta (3)
		\end{align}
		\begin{align}
		V_4(r)=\frac{L_{\gamma}}{r} \left(L_{\gamma}^3+2 \pi ^2 L_{\gamma}+64 \zeta (3)\right)+\frac{19 \pi ^4}{15 r}
		\end{align}
		\par
		Restricted Fourier transform contains hypergeometric functions with array of $\frac{1}{2}$ in first argument and $\frac{3}{2}$ in the second argument and if we define:
		\begin{align}
		\text{Si}(r \mu_f)\equiv \left(\mu_f r\right)\times H_1 , \quad \quad
		_nF_{n+1}\big(\frac{1}{2},\frac{1}{2},\dots;\frac{3}{2},\frac{3}{2},\frac{3}{2},\dots;-\frac{1}{4} r^2 \mu _f^2\big)\equiv H_{n}
		\label{hyp}
		\end{align}
		then the uncontrolled contribution to static potential without the ultrasoft term is given by:
		\begin{align}
			\delta V(r,\mu_f)=\frac{2 \mu_f}{\pi}\sum_{i=0}^{4} \sum_{j=0}^{i}x^{i+1} \delta V_j(r,\mu_f)+\order{x^6}
		\end{align}
		where, 
		\begin{align}
			\delta V_0(r,\mu_f)&=H_1\,,\quad 
			\delta V_1(r,\mu_f)=2 H_2-H_1 L_{\mu_f} \quad \delta V_2(r,\mu_f)=-4 H_2 L_{\mu_f}+\frac{1}{3} H_1 \left(3 L_{\mu_f}^2-\pi ^2\right)+8 H_3\nn\\\delta V_3(r,\mu_f)&=-24 H_3 L_{\mu_f}+H_2 \left(6 L_{\mu_f}^2-2 \pi ^2\right)+H_1 \left(-L_{\mu_f}^3+\pi ^2 L_{\mu_f}-16 \zeta (3)\right)+48 H_4
			\nn \\\delta V_4(r,\mu_f)&=\frac{1}{5} H_1 \left(5 L_{\mu_f}^4-10 \pi ^2 L_{\mu_f}^2+320 \zeta (3) L_{\mu_f}-3 \pi ^4\right)+\frac{1}{5} H_2 \left(-40 L_{\mu_f}^3+40 \pi ^2 L_{\mu_f}-640 \zeta (3)\right)\nn\\&+\frac{1}{5} H_3 \left(240 L_{\mu_f}^2-80 \pi ^2\right)-192 H_4 L_{\mu_f}+384 H_5
		\end{align}
		\section{Restricted Version of the Static Energy in Position Space\label{app:resVr}}

		Uncontrolled contribution to the static energy in RFT scheme is given by:
		\begin{align}
		\delta E&(r,\mu,\mu_f)=-\left(2 C_F H_1 \mu_f x \right)\Bigg(1+x \big(2 \tilde{H}_2 T_{1,1,0}+T_{1,1,0} L_{\mu_f}+T_{1,0,0}\big)+x^2 \big(4 \tilde{H}_2 T_{2,2,0} L_{\mu_f}+2 \tilde{H}_2 T_{2,1,0}+8 \tilde{H}_3 T_{2,2,0}+T_{2,0,0}\nonumber\\&+T_{2,2,0} L_{\mu_f}^2+T_{2,1,0} L_{\mu_f}\big) +x^3 \Big(T_{3,0,0}+2 \tilde{H}_2 T_{3,1,0}+8 \tilde{H}_3 T_{3,2,0}+48 \tilde{H}_4 T_{3,3,0}+L_{\mu_f}^2 \big(6 \tilde{H}_2 T_{3,3,0}+T_{3,2,0}\big)+T_{3,3,0} L_{\mu_f}^3\nonumber\\&+L_{\mu_f} \big(4 \tilde{H}_2 T_{3,2,0}+24 \tilde{H}_3 T_{3,3,0}+T_{3,1,0}\big) + \frac{1}{144} \pi ^2 C_A^3 \big(12 \tilde{H}_2+12 \log \left(H_1\right)+12 L_{us}-12 \gamma _E -10+24 L_2\big) \Big)\nonumber\\&+x^4\Big(T_{4,0,0}+2 \tilde{H}_2 T_{4,1,0}+8 \tilde{H}_3 T_{4,2,0}+48 \tilde{H}_4 T_{4,3,0}+384 \tilde{H}_5 T_{4,4,0}+L_{\mu_f}^2 \big(6 \tilde{H}_2 T_{4,3,0}+48 \tilde{H}_3 T_{4,4,0}+T_{4,2,0}\big)+T_{4,4,0} L_{\mu_f}^4 \nonumber\\&+L_{\mu_f} \Big(\frac{1}{144} \pi ^2 C_A^3 \big(48 \beta _0 \tilde{H}_2+12 T_{1,1,0}-48 \gamma _E \beta _0-40 \beta _0+96 \beta _0 L_2+48 \beta _0 \log \left(H_1\right)+48 \beta _0 L_{us}\big)+T_{4,1,0}+24 \tilde{H}_3 T_{4,3,0}\nonumber\\&+192 \tilde{H}_4 T_{4,4,0}+4 \tilde{H}_2 T_{4,2,0}\Big)+L_{\mu_f}^3 \big(8 \tilde{H}_2 T_{4,4,0}+T_{4,3,0}\big)+\frac{\pi ^4 C_A^3 }{144} \Big(12 \tilde{H}_2-9 \beta _0+12 \log \left(H_1\right)+12 L_{us}-9 \gamma _E -8+18 L_2\nonumber\\&-3 L_{\pi}\Big)+\frac{\pi ^2 C_A^3 }{144} \Big(36 \tilde{H}_2 T_{1,0,0}+24 \tilde{H}_2 T_{1,1,0}-96 \gamma _E \beta _0 \tilde{H}_2-40 \beta _0 \tilde{H}_2+168 \beta _0 \tilde{H}_3+144 \beta _0 L_2 \tilde{H}_2+72 \beta _0 \tilde{H}_2 \log \left(H_1\right)+39 \tilde{H}_2\nonumber\\&+72 \beta _0 L_{us}\tilde{H}_2+36 T_{1,0,0} \log \left(H_1\right)+36 L_{us}T_{1,0,0}-36 \gamma _E T_{1,0,0}-18 T_{1,0,0}+72 L_2 T_{1,0,0}-84+12 \gamma _E ^2 \beta _0-15 \gamma _E \beta _0-\frac{134 \beta _0}{3}\nonumber\\&-48 \beta _0 L_2 ^2+70 \beta _0 L_2-5 \beta _0 L_{\pi}-12 \beta _0 \log ^2\left(H_1\right)-48 \beta _0 L_2 \log \left(H_1\right)+40 \beta _0 \log \left(H_1\right)-24 \beta _0 L_{us}\log \left(H_1\right)+39 \log \left(H_1\right)\nonumber\\&-12 \beta _0 (L_{us})^2+40 \beta _0 L_{us}-48 \beta _0 L_{us}L_2+39 L_{us}+72 \zeta (3)-\frac{117 \gamma _E }{4}+\frac{117 L_2}{2}-\frac{39 L_{\pi}}{4}\Big)\Big) \Bigg)
		\end{align}
		where $L_{us}=\log \left(\frac{ C_A x}{2}\right)$, $L_{\mu_f}=\log \left(\frac{\mu ^2}{\mu_f^2}\right)$, $L_{\pi}=\log(\pi)$, $L_{2}=\log(2)$ and $\tilde{H_i}\equiv H_i / H_1$. Note that $H_i's$ are defined by eq(\ref{hyp}) in appendix~\ref{app:Vr}
		\end{widetext}
	

\begin{thebibliography}{99}
	
		\bibitem{Caswell:1985ui}
		W.~E.~Caswell and G.~P.~Lepage,
		Phys. Lett. B \textbf{167} (1986), 437-442
		\bibitem{Bodwin:1994jh}
		G.~T.~Bodwin, E.~Braaten and G.~P.~Lepage,
		Phys. Rev. D \textbf{51} (1995), 1125-1171
		[arXiv:hep-ph/9407339 [hep-ph]].
		\bibitem{Pineda:1997bj}
		A.~Pineda and J.~Soto,
		Nucl. Phys. B Proc. Suppl. \textbf{64} (1998), 428-432
		[arXiv:hep-ph/9707481 [hep-ph]].
		\bibitem{Brambilla:2004jw}
		N.~Brambilla, A.~Pineda, J.~Soto and A.~Vairo,
		Rev.\ Mod.\ Phys.\ {\bf 77} (2005) 1423
		[hep-ph/0410047].
		\bibitem{Anzai:2009tm}
		C.~Anzai, Y.~Kiyo and Y.~Sumino,
		Phys.\ Rev.\ Lett.\ {\bf 104} (2010) 112003
		[arXiv:0911.4335 [hep-ph]].
		\bibitem{Smirnov:2009fh}
		A.~V.~Smirnov, V.~A.~Smirnov and M.~Steinhauser,
		Phys.\ Rev.\ Lett.\ {\bf 104} (2010) 112002
		[arXiv:0911.4742 [hep-ph]].
		\bibitem{Lee:2016cgz} 
		R.~N.~Lee, A.~V.~Smirnov, V.~A.~Smirnov and M.~Steinhauser,
		Phys.\ Rev.\ D {\bf 94}, no. 5, 054029 (2016),
		[arXiv:1608.02603 [hep-ph]].
		\bibitem{Appelquist:1977es}
		T.~Appelquist, M.~Dine and I.~J.~Muzinich,
		Phys.\ Rev.\ D {\bf 17} (1978) 2074.
		\bibitem{Brambilla:1999qa}
		N.~Brambilla, A.~Pineda, J.~Soto and A.~Vairo,
		Phys.\ Rev.\ D {\bf 60} (1999) 091502
		[hep-ph/9903355].
%
		\bibitem{Pineda:2000gza}
		A.~Pineda and J.~Soto,
		Phys.\ Lett.\ B {\bf 495} (2000) 323
		[hep-ph/0007197].
%
		\bibitem{Brambilla:2006wp}
		N.~Brambilla, X.~Garcia i Tormo, J.~Soto and A.~Vairo,
		Phys.\ Lett.\ B {\bf 647} (2007) 185
		[hep-ph/0610143].
		\bibitem{Brambilla:2009bi}
		N.~Brambilla, A.~Vairo, X.~Garcia i Tormo and J.~Soto,
		Phys.\ Rev.\ D {\bf 80} (2009) 034016
		[arXiv:0906.1390 [hep-ph]].
%
		\bibitem{Bazavov:2014soa} 
		A.~Bazavov, N.~Brambilla, X.~Garcia i Tormo, P.~Petreczky, J.~Soto and A.~Vairo,
		Phys.\ Rev.\ D {\bf 90}, no. 7, 074038 (2014),
		[arXiv:1407.8437 [hep-ph]].
%
		\bibitem{Karbstein:2014bsa}
		F.~Karbstein, A.~Peters and M.~Wagner,
		JHEP {\bf 1409} (2014) 114
		[arXiv:1407.7503 [hep-ph]].
%
		\bibitem{Karbstein:2018mzo}
		F.~Karbstein, M.~Wagner and M.~Weber,
		Phys.\ Rev.\ D {\bf 98} (2018) no.11, 114506
		[arXiv:1804.10909 [hep-ph]].
%
		\bibitem{Takaura:2018vcy}
		H.~Takaura, T.~Kaneko, Y.~Kiyo and Y.~Sumino,
		JHEP {\bf 1904} (2019) 155
		[arXiv:1808.01643 [hep-ph]].
%
		\bibitem{Takaura:2018lpw}
		H.~Takaura, T.~Kaneko, Y.~Kiyo and Y.~Sumino,
		Phys.\ Lett.\ B {\bf 789} (2019) 598
		[arXiv:1808.01632 [hep-ph]].
%
		\bibitem{Bazavov:2019qoo}
		A.~Bazavov \textit{et al.} [TUMQCD],
		Phys. Rev. D \textbf{100} (2019) no.11, 114511
		[arXiv:1907.11747 [hep-lat]].
%
%
%
		\bibitem{Brambilla:1999xf}
		N.~Brambilla, A.~Pineda, J.~Soto and A.~Vairo,
		Nucl. Phys. B \textbf{566} (2000), 275
		[arXiv:hep-ph/9907240 [hep-ph]].
		\bibitem{Penin:2014zaa}
		A.~A.~Penin and N.~Zerf,
		JHEP {\bf 1404} (2014) 120
		[arXiv:1401.7035 [hep-ph]].
		\bibitem{Pineda:2001zq} 
		A.~Pineda,
		JHEP {\bf 0106}, 022 (2001)
		[hep-ph/0105008].
		\bibitem{Ayala:2014yxa}
		C.~Ayala, G.~Cvetič and A.~Pineda,
		JHEP {\bf 1409} (2014) 045
		[arXiv:1407.2128 [hep-ph]].
		\bibitem{Beneke:2014pta}
		M.~Beneke, A.~Maier, J.~Piclum and T.~Rauh,
		Nucl.\ Phys.\ B {\bf 891} (2015) 42
		[arXiv:1411.3132 [hep-ph]].
		\bibitem{Kiyo:2015ufa}
		Y.~Kiyo, G.~Mishima and Y.~Sumino,
		Phys.\ Lett.\ B {\bf 752} (2016) 122
		Erratum: [Phys.\ Lett.\ B {\bf 772} (2017) 878]
		[arXiv:1510.07072 [hep-ph]].
%
		\bibitem{Beneke:2015zqa}
		M.~Beneke and M.~Steinhauser,
		Nucl.\ Part.\ Phys.\ Proc.\ {\bf 261-262} (2015) 378
		[arXiv:1506.07962 [hep-ph]].
%
		\bibitem{Kiyo:2015ooa}
		Y.~Kiyo, G.~Mishima and Y.~Sumino,
		JHEP {\bf 1511} (2015) 084
		[arXiv:1506.06542 [hep-ph]].
%
		\bibitem{Mateu:2017hlz}
		V.~Mateu and P.~G.~Ortega,
		JHEP {\bf 1801} (2018) 122
		[arXiv:1711.05755 [hep-ph]].
%
		\bibitem{Peset:2018ria}
		C.~Peset, A.~Pineda and J.~Segovia,
		JHEP {\bf 1809} (2018) 167
		[arXiv:1806.05197 [hep-ph]].
%
		\bibitem{Hoang:2000fm}
		A.~H.~Hoang,
		hep-ph/0008102.
%
%
		\bibitem{Hoang:2013uda}
		A.~H.~Hoang and M.~Stahlhofen,
		JHEP {\bf 1405} (2014) 121
		[arXiv:1309.6323 [hep-ph]].
%
%
		\bibitem{Beneke:2013jia}
		M.~Beneke, Y.~Kiyo and K.~Schuller,
		arXiv:1312.4791 [hep-ph].
%
		\bibitem{Beneke:2015kwa}
		M.~Beneke, Y.~Kiyo, P.~Marquard, A.~Penin, J.~Piclum and M.~Steinhauser,
		Phys.\ Rev.\ Lett.\ {\bf 115} (2015) no.19, 192001
		[arXiv:1506.06864 [hep-ph]].
		\bibitem{Chishtie:2001mf} 
		F.~A.~Chishtie and V.~Elias,
		Phys.\ Lett.\ B {\bf 521}, 434 (2001)
		[hep-ph/0107052].
		\bibitem{Maxwell:1999dv} 
		C.~J.~Maxwell,
		Nucl.\ Phys.\ Proc.\ Suppl.\ {\bf 86}, 74 (2000)
		[hep-ph/9908463].
		\bibitem{Maxwell:2000mm} 
		C.~J.~Maxwell and A.~Mirjalili,
		Nucl.\ Phys.\ B {\bf 577}, 209 (2000)
		[hep-ph/0002204].
		\bibitem{Maxwell:2001he} 
		C.~J.~Maxwell and A.~Mirjalili,
		Nucl.\ Phys.\ B {\bf 611}, 423 (2001)
		[hep-ph/0103164].
		\bibitem{Ahmady:1999xg} 
		M.~R.~Ahmady, F.~A.~Chishtie, V.~Elias and T.~G.~Steele,
		Phys.\ Lett.\ B {\bf 479}, 201 (2000)
		[hep-ph/9910551].
		\bibitem{Ahmady:2002fd} 
		M.~R.~Ahmady, F.~A.~Chishtie, V.~Elias, A.~H.~Fariborz, N.~Fattahi, D.~G.~C.~McKeon, T.~N.~Sherry and T.~G.~Steele,
		Phys.\ Rev.\ D {\bf 66}, 014010 (2002)
		[hep-ph/0203183].
		\bibitem{Ahmady:2002pa} 
		M.~R.~Ahmady, F.~A.~Chishtie, V.~Elias, A.~H.~Fariborz, D.~G.~C.~McKeon, T.~N.~Sherry, A.~Squires and T.~G.~Steele,
		Phys.\ Rev.\ D {\bf 67}, 034017 (2003)
		[hep-ph/0208025].
		\bibitem{Abbas:2012py}
		G.~Abbas, B.~Ananthanarayan and I.~Caprini,
		Phys.\ Rev.\ D {\bf 85} (2012) 094018
		[arXiv:1202.2672 [hep-ph]].
		\bibitem{Ananthanarayan:2016kll} 
		B.~Ananthanarayan and D.~Das,
		Phys.\ Rev.\ D {\bf 94}, no. 11, 116014 (2016)
		[arXiv:1610.08900 [hep-ph]].
		\bibitem{Karbstein:2013zxa}
		F.~Karbstein,
		JHEP {\bf 1404} (2014) 144
		[arXiv:1311.7351 [hep-ph]].
		\bibitem{Beneke:1997zp}
		M.~Beneke and V.~A.~Smirnov,
		Nucl. Phys. B \textbf{522} (1998), 321-344
		doi:10.1016/S0550-3213(98)00138-2
		[arXiv:hep-ph/9711391 [hep-ph]].
		\bibitem{Appelquist:1977tw}
		T.~Appelquist, M.~Dine and I.~J.~Muzinich,
		Phys.\ Lett.\ {\bf 69B} (1977) 231.
		\bibitem{Susskind} 
		L.~Susskind,
		\emph{Coarse grained quantum chromodynamics} in R. Balian and C. H.
		Llewellyn Smith (eds.), \emph{Weak and electromagnetic interactions at high energy} (North Holland, Amsterdam, 1977).		
		\bibitem{Fischler:1977yf} 
		W.~Fischler,
		Nucl.\ Phys.\ B {\bf 129}, 157 (1977).
		\bibitem{Billoire:1979ih} 
		A.~Billoire,
		Phys.\ Lett.\ {\bf 92B}, 343 (1980).
		\bibitem{Melles:1998dj} 
		M.~Melles,
		Phys.\ Rev.\ D {\bf 58}, 114004 (1998)
		[hep-ph/9805216].
		\bibitem{Schroder:1998vy} 
		Y.~Schroder,
		Phys.\ Lett.\ B {\bf 447}, 321 (1999)
		[hep-ph/9812205].		
		\bibitem{peter:1998ml}
		M.~Peter,
		Phys.\ Rev.\ Lett.\ {\bf 78} (1997) 602, Nucl.\ Phys.\ B {\bf 501} (1997) 471 [hep-ph/9610209];
		\bibitem{Peter:1997me}
		M.~Peter,
		Nucl.\ Phys.\ B {\bf 501} (1997) 471
		[hep-ph/9702245].
		\bibitem{Melles:2000ml}
		M.~Melles,
		Phys.\ Rev.\ D {\bf 62} (2000) 074019
		\bibitem{Melles:2000ey}
		M.~Melles,
		Nucl.\ Phys.\ Proc.\ Suppl.\ {\bf 96} (2001) 472
		[hep-ph/0009085].
		\bibitem{sumino:2002ms}
		S.~Recksiegel and Y.~Sumino,
		Phys.\ Rev.\ D {\bf 65} (2002) 054018
		[hep-ph/0109122].
		\bibitem{Gross:1973id}
		D.~J.~Gross and F.~Wilczek,
		Phys.\ Rev.\ Lett.\ {\bf 30} (1973) 1343.
		\bibitem{Caswell:1974gg}
		W.~E.~Caswell,
		Phys.\ Rev.\ Lett.\ {\bf 33} (1974) 244.
		\bibitem{Jones:1974mm}
		D.~R.~T.~Jones,
		Nucl.\ Phys.\ B {\bf 75} (1974) 531.
		\bibitem{Tarasov:1980au}
		O.~V.~Tarasov, A.~A.~Vladimirov and A.~Y.~Zharkov,
		Phys.\ Lett.\ {\bf 93B} (1980) 429.
		\bibitem{Larin:1993tp}
		S.~A.~Larin and J.~A.~M.~Vermaseren,
		Phys.\ Lett.\ B {\bf 303} (1993) 334
		[hep-ph/9302208].
		\bibitem{vanRitbergen:1997va}
		T.~van Ritbergen, J.~A.~M.~Vermaseren and S.~A.~Larin,
		Phys.\ Lett.\ B {\bf 400} (1997) 379
		[hep-ph/9701390].
		\bibitem{Czakon:2004bu}
		M.~Czakon,
		Nucl.\ Phys.\ B {\bf 710} (2005) 485
		[hep-ph/0411261].
		\bibitem{Baikov:2016tgj}
		P.~A.~Baikov, K.~G.~Chetyrkin and J.~H.~Kühn,
		Phys.\ Rev.\ Lett.\ {\bf 118} (2017) no.8, 082002
		[arXiv:1606.08659 [hep-ph]].
		\bibitem{Herzog:2017ohr}
		F.~Herzog, B.~Ruijl, T.~Ueda, J.~Vermaseren and A.~Vogt,
		JHEP \textbf{02} (2017), 090
		[arXiv:1701.01404 [hep-ph]].
		\bibitem{Elias:1998bi} 
		V.~Elias, T.~G.~Steele, F.~Chishtie, R.~Migneron and K.~B.~Sprague,
		Phys.\ Rev.\ D {\bf 58}, 116007 (1998)
		[hep-ph/9806324].
		\bibitem{Chishtie:1998rz} 
		F.~Chishtie, V.~Elias and T.~G.~Steele,
		Phys.\ Rev.\ D {\bf 59}, 105013 (1999)
		[hep-ph/9812498].
		\bibitem{Chishtie:2000ex} 
		F.~A.~Chishtie and V.~Elias,
		Phys.\ Lett.\ B {\bf 499}, 270 (2001)
		[hep-ph/0008319].
		\bibitem{Elias:2000iw} 
		V.~Elias, F.~A.~Chishtie and T.~G.~Steele,
		J.\ Phys.\ G {\bf 26}, 1239 (2000)
		[hep-ph/0004140].
		\bibitem{Ellis:1996zn}
		J.~R.~Ellis, M.~Karliner and M.~A.~Samuel,
		Phys.\ Lett.\ B {\bf 400} (1997) 176
		[hep-ph/9612202].
		\bibitem{Ellis:1997sb} 
		J.~R.~Ellis, I.~Jack, D.~R.~T.~Jones, M.~Karliner and M.~A.~Samuel,
		Phys.\ Rev.\ D {\bf 57}, 2665 (1998)
		[hep-ph/9710302].
		\bibitem{Samuel:1995jc}
		M.~A.~Samuel, J.~R.~Ellis and M.~Karliner,
		Phys.\ Rev.\ Lett.\ {\bf 74} (1995) 4380
		[hep-ph/9503411].
		\bibitem{Jezabek:1998wk}
		M.~Jezabek, M.~Peter and Y.~Sumino,
		Phys.\ Lett.\ B {\bf 428} (1998) 352
		[hep-ph/9803337].
	\end{thebibliography}
\end{document}